\pdfoutput=1
\documentclass[twocolumn]{aastex62}

\newcommand*\dif{\mathop{}\!\mathrm{d}}

\received{November 25, 2018}
\revised{January 28, 2019}
\accepted{January 29, 2019}
\submitjournal{ApJ}

\shorttitle{Effect of Mass-Draining on a Line Current}
\shortauthors{Jenkins et al.}
\usepackage{amsmath}
\allowdisplaybreaks
\usepackage{amssymb}
\usepackage{mathpazo}
\usepackage{graphicx}
\usepackage{textcomp}

\usepackage{epsfig}
\usepackage{times}
\usepackage{natbib}
\usepackage{url}
\usepackage{color}
\usepackage{gensymb}
\usepackage{comment}

\definecolor{ao(english)}{rgb}{0.0, 0.5, 0.0} % darkgreen definition  

\newcommand{\BE}{\begin{equation}}
\newcommand{\EE}{\end{equation}}
\newcommand{\BA}{\begin{eqnarray}}
\newcommand{\EA}{\end{eqnarray}}

 \newcommand{\eq}[1]{Eq.~(\ref{eq_#1})}

\newcommand{\Int}[2]{\ensuremath{\mathchoice%
        {\displaystyle\int_{#1}^{#2}}
        {\displaystyle\int_{#1}^{#2}}
        {\int_{#1}^{#2}}
        {\int_{#1}^{#2}}
        }}

\newcommand{\rmd}{{\rm d }}

\newcommand{\Be}{B_{\rm ext}}
\newcommand{\hm}{h_{\rm m}}
\newcommand{\ho}{h_{\rm 0}}
\newcommand{\Ie}[1]{I_{\rm eq,#1}}
\renewcommand{\Im}{I_{\rm m}}
\newcommand{\Io}{I_{\rm 0}}
\newcommand{\ncrit}{n_{\rm crit}}

\usepackage{tabularx}

\begin{document}

\title{MODELLING THE EFFECT OF MASS-DRAINING ON PROMINENCE ERUPTIONS}

\author[0000-0002-8975-812X]{Jack M. Jenkins}
\affiliation{Mullard Space Science Laboratory, University College London, Dorking, RH5 6NT, UK; \href{mailto:jack.jenkins.16@ucl.ac.uk}{jack.jenkins.16@ucl.ac.uk}}

\author[0000-0002-2071-9405]{Matthew Hopwood}
\affiliation{School of Mathematics, University of Birmingham, Birmingham B15 2TT, UK}
\affiliation{School of Mathematical Sciences, The University of Adelaide, Adelaide, SA 5005, Australia}

\author[0000-0001-8215-6532]{Pascal D{\'e}moulin}
\affiliation{LESIA-Observatoire de Paris, CNRS, UPMC Univ Paris 06, Univ. Paris-Diderot, Meudon Cedex, France}

\author[0000-0001-7809-0067]{Gherardo Valori}
\affiliation{Mullard Space Science Laboratory, University College London, Dorking, RH5 6NT, UK; \href{mailto:jack.jenkins.16@ucl.ac.uk}{jack.jenkins.16@ucl.ac.uk}}

\author[0000-0001-5810-1566]{Guillaume Aulanier}
\affiliation{LESIA-Observatoire de Paris, CNRS, UPMC Univ Paris 06, Univ. Paris-Diderot, Meudon Cedex, France}

\author[0000-0003-3137-0277]{David M. Long}
\affiliation{Mullard Space Science Laboratory, University College London, Dorking, RH5 6NT, UK; \href{mailto:jack.jenkins.16@ucl.ac.uk}{jack.jenkins.16@ucl.ac.uk}}

\author[0000-0002-2943-5978]{Lidia van Driel-Gesztelyi}
\affiliation{Mullard Space Science Laboratory, University College London, Dorking, RH5 6NT, UK; \href{mailto:jack.jenkins.16@ucl.ac.uk}{jack.jenkins.16@ucl.ac.uk}}
\affiliation{LESIA-Observatoire de Paris, CNRS, UPMC Univ Paris 06, Univ. Paris-Diderot, Meudon Cedex, France}
\affiliation{Konkoly Observatory of the Hungarian Academy of Sciences, Budapest, Hungary}

%% Note that the \and command from previous versions of AASTeX is now
%% depreciated in this version as it is no longer necessary. AASTeX 
%% automatically takes care of all commas and "and"s between authors names.

%% AASTeX 6.2 has the new \collaboration and \nocollaboration commands to
%% provide the collaboration status of a group of authors. These commands 
%% can be used either before or after the list of corresponding authors. The
%% argument for \collaboration is the collaboration identifier. Authors are
%% encouraged to surround collaboration identifiers with ()s. The 
%% \nocollaboration command takes no argument and exists to indicate that
%% the nearby authors are not part of surrounding collaborations.

%% Mark off the abstract in the ``abstract'' environment. 
\begin{abstract}

Quiescent solar prominences are observed to exist within the solar atmosphere for up to several solar rotations. Their eruption is commonly preceded by a slow increase in height that can last from hours to days. This increase in the prominence height is believed to be due to their host magnetic flux rope transitioning through a series of neighbouring quasi-equilibria before the main loss-of-equilibrium that drives the eruption. Recent work suggests that the removal of prominence mass from a stable, quiescent flux rope is one possible cause for this change in height. However, these conclusions are drawn from observations and are subject to interpretation. Here we present a simple model to quantify the effect of ``mass-draining'' during the pre-eruptive height-evolution of a solar flux rope. The flux rope is modeled as a line current suspended within a background potential magnetic field. We first show that the inclusion of mass, up to $10^{12}$~kg, can modify the height at which the line current experiences loss-of-equilibrium by up to 14\%. Next, we show that the rapid removal of mass prior to the loss-of-equilibrium can allow the height of the flux rope to increase sharply and without upper bound as it approaches its loss-of-equilibrium point. This indicates that the critical height for the loss-of-equilibrium can occur at a range of heights depending explicitly on the amount and evolution of mass within the flux rope. Finally, we demonstrate that for the same amount of drained mass, the effect on the height of the flux rope is up to two order of magnitude larger for quiescent than for active region prominences.
 
\end{abstract}

%% Keywords should appear after the \end{abstract} command. 
%% See the online documentation for the full list of available subject
%% keywords and the rules for their use.

%% From the front matter, we move on to the body of the paper.
%% Sections are demarcated by \section and \subsection, respectively.
%% Observe the use of the LaTeX \label
%% command after the \subsection to give a symbolic KEY to the
%% subsection for cross-referencing in a \ref command.
%% You can use LaTeX's \ref and \label commands to keep track of
%% cross-references to sections, equations, tables, and figures.
%% That way, if you change the order of any elements, LaTeX will
%% automatically renumber them.
%%
%% We recommend that authors also use the natbib \citep
%% and \citet commands to identify citations.  The citations are
%% tied to the reference list via symbolic KEYs. The KEY corresponds
%% to the KEY in the \bibitem in the reference list below. 

\section{Introduction} \label{sec:intro}

Coronal mass ejections (CMEs) are complex bundles of magnetic field and material that erupt from the solar atmosphere out into the heliosphere. A key feature often measured within their interplanetary counterpart is a rotation of the magnetic field vector as spacecraft cross the magnetic structure, a property believed to be indicative of a magnetic flux rope \citep[\textit{e.g.},][]{Burlaga:1988,Palmerio:2017,James:2017}. In addition, the existence of a flux rope in the solar atmosphere has often been related to the formation of filament systems; elongated structures observed in absorption on the solar disk \citep{Priest:1989,Aulanier:1998b}. Filaments are interpreted as strands of dense material suspended in the low-coronal atmosphere. Such structures are historically identified as prominences when observed above the limb, and we shall henceforth use the term prominence to describe these structures, unless otherwise indicated. The observational signature of the on-disk counterpart, a filament, provides no immediate evidence for the suspended nature of the material above the solar surface \citep{vanballegooijen:1989,Martin:1998,Gibson:2006,Regnier:2011}. Prominences have been observed for up to several solar rotations, occasionally within a coronal cavity when a prominence quasi-parallel to the equator is projected above the limb. A pre-eruptive flux rope has been suggested to exist in equilibrium for equally extended periods of time \citep{Rust:2003, Gibson:2004}. 

%The point at which the suggested flux rope forms remains a topic of debate within the solar physics community. The existence of a pre-eruptive magnetic flux rope was first suggested by \citet{Kurokawa:1987}, although its role within solar eruptions has only been studied in detail within the last two decades \citep[\textit{e.g.}][]{Aulanier:1998,Titov:1999,Rust:2003}.

Despite being typically stable features within the solar atmosphere, the final stages of a prominence's life are highly dynamic; the suspended plasma either drains back to the chromosphere, or is ejected into the heliosphere as the core of a CME, or some combination of both \citep{Dere:1997,Schmahl:1977,Regnier:2011}. In the eruptive case, the sudden destabilisation of these structures is also indicative of the destabilisation of the host flux rope. The exact causes for the loss-of-stability of a flux rope are understood to depend on the conditions under which the flux rope formed and the recent evolution of the surrounding magnetic field \citep{Moore:2001,Lynch:2004,Torok:2005,Fan:2007}. Unfortunately, flux ropes are not directly observable in the solar atmosphere as they are magnetic in nature and instrumentation sensitive enough to accurately measure the coronal magnetic field does not yet exist \citep[although preliminary attempts are being made, \textit{e.g.,}][]{Bak-Steslicka:2013, Fan:2018b}. Therefore, in order to effectively study the stability criteria of flux ropes, a combination of observations \citep[\textit{e.g.,}][]{Zuccarello:2014,Zuccarello:2016}, extrapolations \citep[\textit{e.g.,}][]{James:2018}, and simulations \citep[\textit{e.g.,}][]{Fan:2017} are typically used \citep[see also][and references therein]{Cheng:2017}. The simulations are often employed to study the cause of the loss-of-stability of a flux rope in the lead-up to its eruption, with the observations and extrapolations separately offering information about the pre-eruptive configuration.

Before the advent of advanced simulations, early work by \citet{vanTend:1978} presented a 2D analytical model in which the flux rope was approximated as a straight line current suspended at equilibrium in a background potential magnetic field. Although a simplified setup was employed, the authors qualitatively demonstrated that increasing the magnitude of the line current causes its height above the solar surface to increase. This relationship between the current and height of the line current can be represented with an equilibrium curve. In addition, they concluded that there is a point at which an increase in the strength of the line current would no longer result in a solution on the equilibrium curve. At this time, the line current was said to have experienced `loss-of-equilibrium'. Extensions to this model were employed to quantitatively study the balance of forces involved with prominences, and the evolution of this balance prior to an eruption \citep[\textit{e.g.},][]{Low:1981,Demoulin:1988,Martens:1989,Demoulin:1991,Forbes:1991}. However, authors such as \citet{Martens:1989} and \citet{Demoulin:1991} noted that the influence of the gravity term was negligible assuming ``typical'' values for prominence mass, and was unlikely to be able to perturb the equilibrium dominated by the magnetic pressure and tension forces.

More recently, work has been carried out to take this simple line-current approach further and formulate more complex, time-dependent magnetohydrodynamic (MHD) simulations \citep[for a more complete review on the state of these MHD simulations, see][and references therein]{Cheng:2017}. These models contain more physically realistic initial and boundary conditions that allow the construction, evolution, and analysis of a fully 3D flux rope. Importantly, the modern simulations have aligned with the conclusions of authors such as \citet{Martens:1989} and \citet{Demoulin:1991} that the evolution of the magnetic field in and around a flux rope is assumed to be solely responsible for its evolution in time \citep{Demoulin:1998}. Specifically, this low-beta approximation assumes that the pressure and mass of prominence plasma suspended by a flux rope is negligible in comparison with the magnetic pressure and tension forces of the flux rope and its surroundings \citep{Titov:1999,Filippov:2018}. Indeed, this assumption is featured frequently in three decades of modern research. 

However, novel observations and hydrostatic modeling are beginning to suggest that mass may be able to influence the local and global properties of magnetic flux ropes \citep{Low:2003,Petrie:2007,Seaton:2011,Gunar:2013,Bi:2014,Reva:2017,Jenkins:2018}. In particular, the Shafranov shift as explored in \citet{Blokland:2011} details how varying the gravity term in their 2D magnetohydrostatic (MHS) model can cause the axis of their flux rope to decrease the height. Then, the mass-unloading theory \citep[\textit{e.g.,}][]{Low:1999,Forbes:2000,Klimchuk:2001} has been suggested as one possible cause for the eruption of prominences. In this theory, a particularly heavy prominence suddenly unloads all of its mass, reducing the gravitational force acting on the host flux rope and causing it to \textit{spring off} into space as an eruption.

The study of the role of mass evolution within prominence eruptions has typically been isolated to a handful of observational case studies. \citet{Seaton:2011} presented stereoscopic observations of a prominence erupting from an active region in which plasma was observed to unload from the prominence prior to its expansion in height. The authors concluded that in the absence of additional, contrary evidence, these observations were an example of a `mass-unloading' eruption driver. 

Recently, \citet{Jenkins:2018} also presented stereoscopic observations of a quiescent prominence's partial eruption in which `mass-draining' was suggested to have been responsible for the accelerated expansion of the erupting magnetic flux rope. In this case the drained mass does not ultimately drive an eruption, it simply modifies the balance of forces acting on the prominence to a non-negligible degree \citep[see also][]{Reva:2017}. Their conclusion that the mass-draining accelerated the eruption was reached through a quantitative estimation based on the Lorentz force equation, specifically the ratio between the modification of the gravitational force due to the reduction in mass and the force of the background magnetic tension restricting the height-evolution of the flux rope. However, this order of magnitude estimate to the importance of the mass-draining does not properly account for the equilibrium conditions of the host flux rope.

Therefore, in this manuscript, we present an extension to the model developed by \citet{vanTend:1978} that enables the study of the role of mass in the evolution of a line current in quasi-equilibrium. Specifically, we first explore how the inclusion of mass can modify the stability criteria for a line current that represents a flux rope suspending a prominence. We then explore how the removal of mass (or ``draining'') from a pre-eruptive line current, can modify the global height of the line current within the solar atmosphere. The general model is described in Sections~\ref{S:Model} and \ref{S:General}, and applied in Section~\ref{S:Results} to a bipolar background potential magnetic field. In Section~\ref{S:Implications}, we further constrain the model with measurements made from the observations presented by both \citet{Seaton:2011} and \citet{Jenkins:2018}. Finally, a discussion and summary are presented in Section~\ref{S:Conclusions}.

%\newpage

\section{Model Concept}
	\label{S:Model}

\begin{figure}
\centerline{\includegraphics[width=0.3\textwidth, clip=, trim=0 330 700 0]{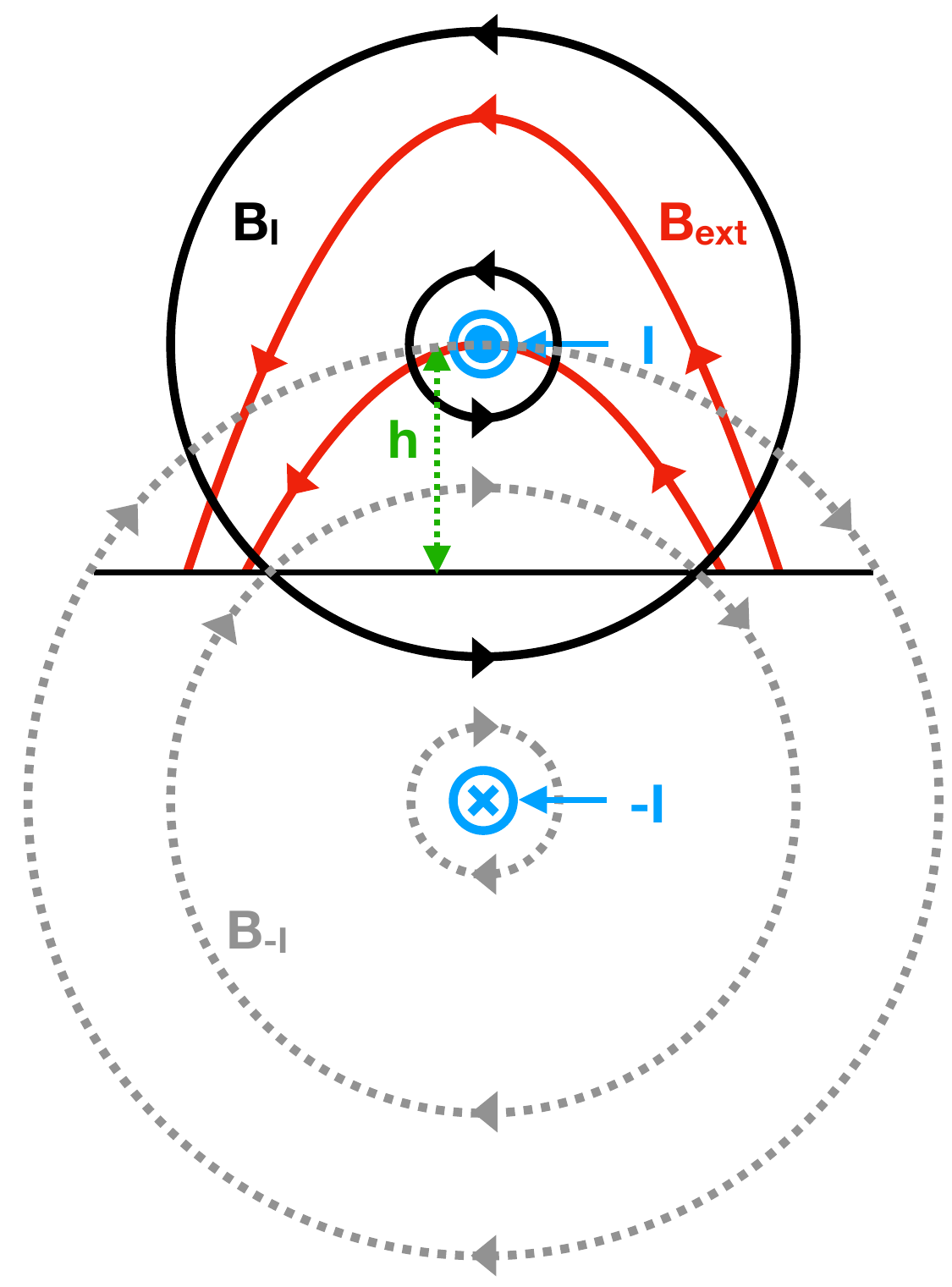}
			}\caption{Cartoon diagram of the model set-up. The inverse magnetic configuration is formed by the superposition of three fields: the external potential field $B_\mathrm{ext}$ (solid-red), and the field generated by the line current (located at z=h) and its image (located at z=-h), drawn with solid-black and dashed-grey lines respectively. The line current at z=h is maintained by the balance of two Lorentz forces, an upward (hoop) force due to the magnetic field generated by the image line current, and a downward force from a stabilizing external potential field $B_\mathrm{ext}$. Model concept is identical to that presented by \citet{vanTend:1978}.}\label{fig:vtend_kuperus}
\end{figure}

Following the formulation outlined in \citet{Demoulin:2010}, hereafter DA10, a flux rope is modeled in cartesian coordinates as a magnetic field generated by an infinitely long, straight line current \textit{I} at a given height \textit{h} above the photosphere. The justification for the choice of a straight line current over a curved line current lies in the assumed property of quiescent prominences being oriented largely horizontal to the surface. The majority of curvature may be assumed to be localised at the footpoints of the host magnetic flux rope that are located far from the center of the prominence. An ``image'' line current \textit{-I} is introduced under the photosphere that runs anti-parallel to the ``real'' line current. Following \citet{vanTend:1978}, the additional image magnetic field beneath the surface results in no modification to the vertical, z, component of the photospheric magnetic field. This ``image'' current acts to increase the height of the ``real'' line current. The straight line current is then added to a background potential magnetic field \textit{$B_\mathrm{ext}$} that acts to force the line current towards the photosphere. A cartoon representation of these different field contributions is shown in Figure~\ref{fig:vtend_kuperus}. The total field has an inverse configuration because of the presence of a flux rope \citep[e.g., similar to the configuration of Figure 1a within][]{Petrie:2007}. The line current is then in equilibrium if the sum of forces is zero,

\begin{equation}
	\sum{f}	= 0 \leftrightarrow f_\mathrm{u}	= f_\mathrm{d}
	\Rightarrow IB_\mathrm{-I}	= IB_\mathrm{ext}    \label{eq_f_balance}
\end{equation}
where $f_\mathrm{u}$ is the sum of the upward magnetic forces, the so-called \textit{hoop force}, $f_\mathrm{d}$ is the sum of the downward magnetic forces, $I$ ($-I$) is the real (image) current, $-B_\mathrm{ext}$ is the horizontal background magnetic field component orthogonal to the current at height $h$, and $B_{-I}$ is the strength of the magnetic field as a consequence of the image line current. The image magnetic field $B_\mathrm{-I}$ is derived from Amp\`ere's law,
\BA
	\oint B_\mathrm{-I} \cdot dl &= \mu_0 I, \nonumber \\ 
	\Rightarrow B_\mathrm{-I}&=\frac{\mu_0 I}{2 \pi R}, \label{eq_b_image}
\EA
where the strength of the magnetic field $B_\mathrm{-I}$ is measured at a point in space that is at a distance/height $R=2h$ away from the line current, and $\mu_0$ is the permeability of free space equal to $4\pi~\times~10^{-7}$ in MKS units.

In order to simulate the existence of a prominence within a flux rope, the \citet{vanTend:1978} model is extended to include mass that is set to exist at the same point as the line current, \textit{i.e.}, at height $h$. The inclusion of mass into the system results in an additional downward force that acts to further anchor the line current. In equilibrium, Eq.~(\ref{eq_f_balance}) becomes,
\begin{equation}
	I B_\mathrm{-I}=IB_\mathrm{ext} + mg, \label{eq_f_balance+m}
\end{equation}
where $m$ is the mass of the suspended plasma per unit length and $g$ is the acceleration due to gravity. $g$ is taken independent of $h$ (since $h\ll r_\odot$, where $r_\odot$ is the solar radius) except where explicitly stated.
%\begin{equation}
%    g=\frac{GM}{(h+r_\odot)}, \label{eq_gravity}
%\end{equation}
%where $G$ is the gravitational constant, $M$ is the mass of the Sun, and $r_\odot$ is the solar radius. 
All quantities in \eq{f_balance+m} are defined positive.

%As the model is static and does not account for dynamics, the removal of mass from the line current to simulate mass-draining is achieved through a comparison between equilibrium points on two equilibrium curves, one curve with the mass and one without. It is this perturbation due to the removal of mass that we focus on here.

\newpage
%%%%%%%%%%%%%%%%%%%%%%%%%%%%%%%%%%%%%%%%%%%%%%%%%%%%%%%
\section{General equations}
	\label{S:General}
%%%%%%%%%%%%%%%%%%%%%%%%%%%%%%%%%%%%%%%%%%%%%%%%%%%%%%%
    \subsection{Equilibrium Current}

   Here we will establish the general form of equations that will be applied to a specific $B_\mathrm{ext}$ in the following sections.
   %First, we will analyse below the force applied to a straight line current, of intensity $I$, of mass $m$, located at height $h$, within the external field horizontal component $\Be$, and with the image current of intensity $-I$ and located at height $-h$. 
The force $f$ on the line current, per unit length, is,
  \BE   \label{eq_f}
  f = \frac{\mu_\mathrm{0}I^2}{4\pi h} -I \Be - m g,
  \EE
where $\Be$ is a function of $h$ (as well as other parameters depending on the selected model).
We set $\Be>0$ so that the external magnetic field creates a force oppositely directed to the hoop force, $\mu_\mathrm{0}I^2/4\pi h$, and an equilibrium exists in the limit $m=0$.

%  {\S}{\bf --- Equilibrium: $\Ie{m}$} \\
The electric current needed for equilibrium is given by solving \eq{f} for $I$ with $f=0$,  
  \BE   \label{eq_Iefull}
  \Ie{m} = \frac{2\pi h \Be}{\mu_\mathrm{0}} \pm \sqrt{\left(\frac{2\pi h \Be}{\mu_\mathrm{0}}\right)^2 + \frac{4\pi}{\mu_\mathrm{0}}m\, g\, h} \, ,
  \EE
where we have added the lower index $m$ to indicate that the equilibrium current depends on the mass.
%The equilibrium current without mass is then simply,
%  \BE   \label{eq_Ie0}
%  \Ie{0} = \frac{4 \pi}{\mu_\mathrm{0}}\, h \Be \, .
%  \EE
  
%  {\S}{\bf --- Two equilibria for $\Ie{m}$ selection} \\
With finite mass, \eq{Iefull} provides two equilibria corresponding to the sign selection in front of the square root.  With a negative sign selected, $\Ie{m}<0$, which implies that both magnetic forces are upward and opposite to the gravity force in \eq{f}. This case has a vanishing current in the limit of a vanishing mass and it does not correspond to a force free equilibrium with a flux rope. Therefore, we consider only the second case with a positive sign in front of the square root of \eq{Iefull}, 
  \BE   \label{eq_Ie}
  \Ie{m} = \frac{2\pi h \Be}{\mu_\mathrm{0}} + \sqrt{\left(\frac{2\pi h \Be}{\mu_\mathrm{0}}\right)^2 + \frac{4\pi}{\mu_\mathrm{0}}\,m\, g\, h} \, ,
  \EE

%  {\S}{\bf --- $\Ie{m}$ behavior near $h=0$} \\
Supposing that $\Be (0)$ is finite, then for small enough $h$ values such that $h \ll (\mu_\mathrm{0}\, m\, g / \pi B_\mathrm{ext}^2)$,
  \BE   \label{eq_Ie_hsmall}
  \Ie{m} \approx \frac{2 \pi h \Be}{\mu_\mathrm{0}} + \sqrt{\frac{4\pi}{\mu_\mathrm{0}}\, m\, g\, h} \, .
  \EE
Then, $\Ie{m}$(h) has a square root dependence with $h$ when $h$ is small enough and $m>0$. This behavior changes to a linear dependence when $m=0$.
 
%  {\S}{\bf --- $\Ie{m}$ behavior at large $h$} \\
With $m=0$, $\Ie{0}(0)=0$, and since $\Be$ typically decreases faster than $1/h$ for large $h$ values, $\Ie{0}(h)=(2\pi/\mu_\mathrm{0})\,h \Be$ will tend towards zero at large heights. This implies that $\Ie{0}(h)$ has a maximum (at least one) between small and large heights. However, if $m>0$, $\Ie{m}(h)$ is dominated by the gravity term at large $h$ values once $\Be$ has sufficiently decreased, then $\Ie{m}(h)\approx \sqrt{ (4\pi/\mu_\mathrm{0})\, m\, g\, h}$ is a growing function of $h$ for constant $g$. At even larger $h$ values, as $g$ is inversely proportional to $(r_\odot + h)^2$, then $\Ie{m}(h)$ will again tend towards zero, even for large mass values. Nevertheless, for low enough $m$ and $h$ values, $\Ie{m}(h)$ will still have a minimum at $h=0$ and large heights, and a maximum (at least one) somewhere in between. It is the response of the line current to mass within this region that we focus on during this study.
%\pc{I wonder that being even more precise will be too heavy}

%===================================================

%===================================================

%------------------------------------------------------

%\newpage
\subsection{Dependence of the Equilibrium current on Mass}

%  {\S}{\bf --- Global effect of mass} \\
We investigate below the effect of $m$ on $\Ie{m}$ keeping all other quantities fixed,  
  \BE  
  \frac{\partial \Ie{m}}{\partial m} 
    = \,g\, h  \left/ \sqrt{\left(h\Be\right)^2 + \frac{\mu_\mathrm{0}}{\pi}\, m\, g\, h}
               \right.
    \geq  0  \, .  \label{eq_dIe_dm}
  \EE
Increasing the mass $m$ requires that the magnitude of the current is increased so as to reach a given height (\textit{i.e.}, to increase the hoop force).

%  {\S}{\bf --- Cases with small mass} \\
Next, supposing $m\, g\, h\ll(\pi/\mu_\mathrm{0})(h \Be)^2$, a first order Taylor expansion of \eq{Ie} provides,
  \BE   \label{eq_Ie_small_m}
  \Ie{m} \approx \Ie{0} +  m\, g\/ /\Be  \, .
  \EE
Then, the equilibrium current is comparatively increased by adding a term proportional to the mass and to $1/\Be$. Since $\Be(h)$ is typically a decreasing function of $h$, this implies that $\Ie{m}$ is increasingly separated from $\Ie{0}$ with height. 
%Since an equilibrium is always present at large $h$ values, it implies that, in this model, a successful ejection is only possible with a complete draining of the mass. This is due to the current 2D cartesian description which is invalid at large heights. Therefore, we will concentrate our study at lower heights where the draining and/or the ideal instability are taking place, leading potentially to an eruption.
 
%\newpage
%%%%%%%%%%%%%%%%%%%%%%%%%%%%%%%%%%%%%%%%%%%%%%%%%%%%%%%
\subsection{Mass-draining}
\label{S:Mass_draining}
%%%%%%%%%%%%%%%%%%%%%%%%%%%%%%%%%%%%%%%%%%%%%%%%%%%%%%%
   Finally, we will analyse the effect of draining prominence mass on the host flux rope's equilibrium height and, possibly, its eruption.  We suppose that the draining is fast enough that there is a negligible evolution, through \textit{e.g.}, diffusion \citep[\textit{e.g.,}][]{vanDriel:2003}, of the vertical component of the photospheric field distribution. This is modelled with the image current and implies that the associated potential field, $\Be$, is unchanged. We suppose also that this short-term evolution is done without reconnection. This implies that the magnetic flux, $F$, passing below the flux rope bottom (located at $z=h-a$, where $a$ is the radius of the flux rope / line current) and the photosphere (at $z=0$) is conserved.

%------------------------------------------------------

%  {\S}{\bf --- Global Description} \\
\citet{vanTend:1978} suggested that a line current would experience loss-of-equilibrium if the current magnitude exceeded the maximum of the $I_\mathrm{eq,0}(h)$ curve, as with a classical electric circuit. DA10 \citep[see also][]{Demoulin:1991,Lin:2002} expanded on this by imposing a short-term MHD evolution with flux conservation to study the loss-of-equilibrium of a flux rope. The hybrid MHD / line current approach uses pseudo-time long-term evolution of model parameters, \textit{e.g.}, photospheric flux density $\phi$ or average coronal twist $T$, to overcome the limitations of the classical approach. The evolution of one of the model parameters in this way allows the construction of a family of constant $F$ curves which describe the short-term evolutions. The intersection of these curves with $I_\mathrm{eq,0}$ details how a flux rope evolves as a function of the evolving parameter.

\begin{figure} % Figure explain draining evolution
\centering
\includegraphics[width=0.5\textwidth, clip=, trim=30 0 0 0]{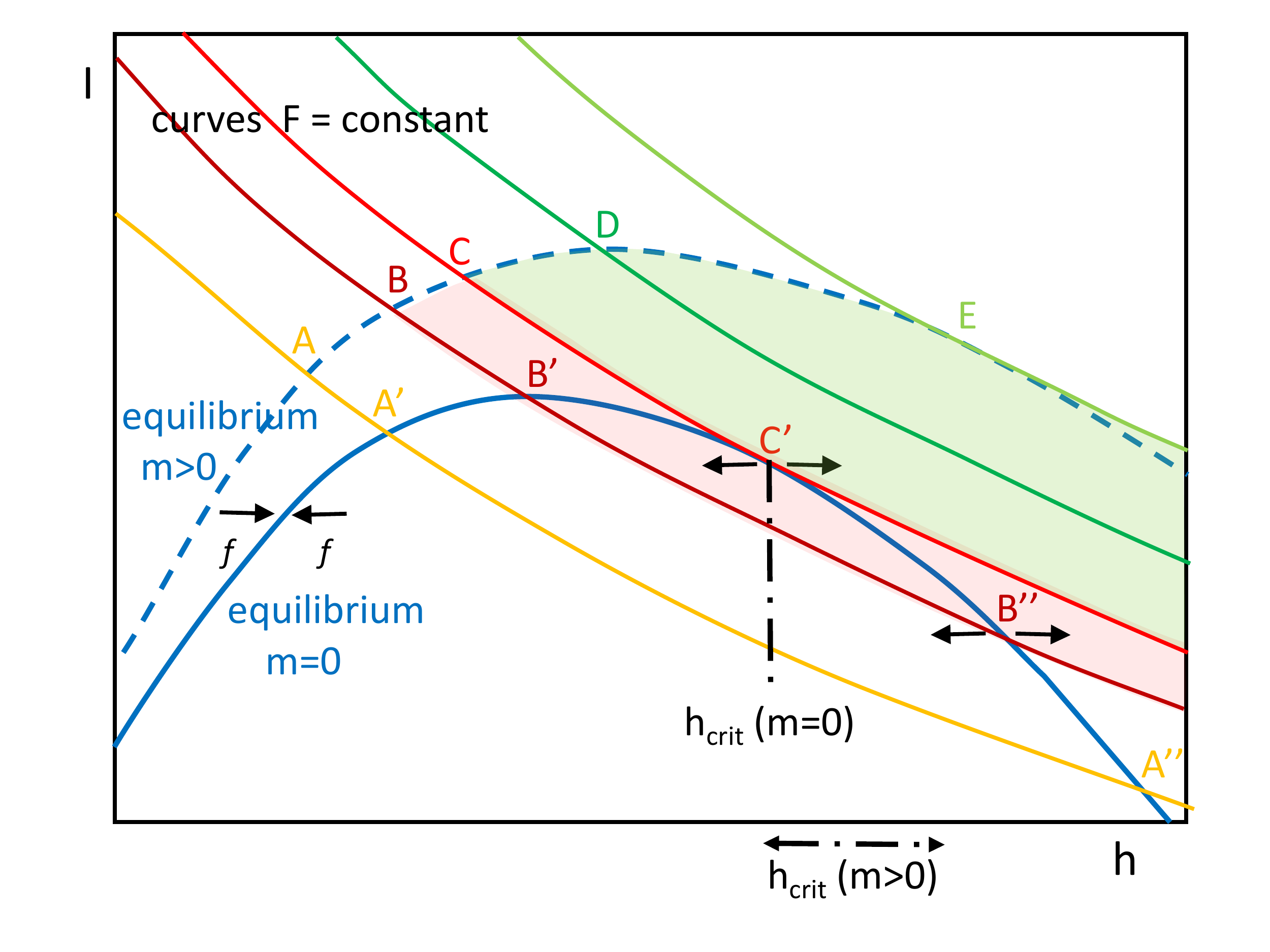}
\caption{Schema showing the possible evolutions with mass-draining.
The equilibrium curve with $m=0$, $\Ie{0}(h)$, is shown with a continuous dark blue line. The equilibrium curve with mass, $\Ie{m}(h)$, is above with a dashed line. The constraint of magnetic flux conservation, \eq{F_conserved}, is shown with the other colored curves representing different starting points along $\Ie{0}(h)$ for draining mass. If the draining mass starts between points C and E, no equilibrium can be reached without mass (region shaded in light green), while if draining is realized before point C (\textit{e.g.}, at point A), another stable equilibrium could be reached. In the region shaded in pink, the finite kinetic energy accumulated may allow the line current to reach the unstable equilibrium without mass (such as point B$''$). The small black arrows indicate the direction of the total force when the line current is slightly shifted away from the equilibrium curve. The critical height(s) $h_\mathrm{crit}$ of the $m=0$ ($m>0$) line current is indicated with the vertical (horizontal) black-dash-dotted lines.
}
\label{fig:draining}
\end{figure}

Two equilibrium curves, $\Ie{m} (h)$ and $\Ie{0} (h)$ are shown in Figure~\ref{fig:draining} together with five curves of fixed magnetic flux $F$, differing from each other as a result of an evolution in, \textit{e.g.}, $B_\mathrm{ext}$. Assuming that we start with a nearly potential coronal configuration, the prominence and its flux rope are supposed to first evolve quasi-statically along the stable equilibrium curve of $\Ie{m} (h)$, with a height growing slowly with time as a result of the evolution of $\Be$.  At some point during this evolution, we have supposed that the draining of the full mass occurs fast enough to keep both $\Be$ and $F$ unchanged, then the evolution is along the corresponding $F=$ constant curve towards larger heights. The general form of the $F=$ constant curve (Eq. (9) of DA10), hereafter defined as $I_\mathrm{evol}(h)$ is,
\BE
    I_\mathrm{evol}(h)=\frac{2}{L_\mathrm{s}}\left( F + \iint\limits_S B_\mathrm{ext}\, \dif y\, \dif z \right), \label{eq_Ievol_general}
    %I_\mathrm{evol}(h)=\frac{F+\frac{2\phi}{\pi}\tan^{-1}\left(\frac{h-a}{D}\right)}{\frac{\mu_\mathrm{0}}{2\pi}\ln\left(\frac{2h}{a}\right)}
\EE
where $L_\mathrm{s}=\frac{\mu_\mathrm{0}\Delta y}{\pi}(\ln(2h/a)+l_\mathrm{i}/2)$, $l_\mathrm{i}$, are the external inductance, and normalised internal inductance, respectively, and $a$ is the radius of the current channel. For this set-up in which the current is focused at the edges of the current channel, $l_\mathrm{i}=0$.

%  {\S}{\bf --- Fix graphically the ideas} \\
The effect of draining the mass depends on the location where it occurs.
  If it occurs at point A of Figure~\ref{fig:draining}, or a nearby one, then a stable equilibrium $\Ie{0}$ exists at the intersection with the $I_\mathrm{evol}$ flux curve (at point A$'$). Comparing the height of stable equilibrium with and without mass linked by the same $I_\mathrm{evol}$ curve, the equilibrium with mass is always at a lower height (\textit{e.g.}, $h_\mathrm{A}<h_\mathrm{A'}$) which is due to the downward gravity force compressing the $B_\mathrm{ext}$ configuration. As the draining point is shifted to larger heights, \textit{e.g.}, at point B, the new equilibrium on $\Ie{0}$ curve is further away, at a larger height, from the initial one on $\Ie{m}$ curve. This is the case until the point C where the $I_\mathrm{evol}$ flux curve only touches the $\Ie{0}$ equilibrium curve tangentially. Equation~(20) of DA10 demonstrates that the equilibrium is linear neutral at this tangent point C$'$, but it is unstable with the non-linear perturbation term taken into account (graphically the $I_\mathrm{evol}$ curve is extending to the right in the region where the force $f$ is pointing towards large $h$ values, so away from the equilibrium curve). 
  
%Therefore, after passing point C, the draining of mass could trigger an eruption. But this is different from a loss-of-equilibrium occurring; the evolution is driven by the draining of mass. If the draining stops early enough, the system could encounter another equilibrium (with lower mass than initially e.g., point B - B$'$). Finally, if point E is reached before any draining occurs, there is a loss-of-equilibrium with an instability bringing the flux rope to larger height.
  
%In summary, starting from a point between C and E, mass-draining can lead to a new equilibrium, with less mass and higher up, or to eruption if too much mass is unloaded.  Then, mass-draining does not necessarily lead to loss-of-equilibrium, it is rather an evolution parameter (like photospheric flux) which drives the system towards a loss-of-equilibrium.
  
%  {\S}{\bf --- Oscillations} \\
After the mass-draining occurs at a point such as A, the total magnetic force will be directed upward, accelerating the flux rope towards the equilibrium curve $\Ie{0}$. However, this equilibrium will be reached with a finite kinetic energy, allowing the line current to continue evolving along the $I_\mathrm{evol}$ curve. The line current will then continue on the other side of the equilibrium point with a change in sign of the total magnetic force. Finally, at some point, the motion will stop and reverse direction leading to an oscillation of the flux rope. This scenario also envisages damped oscillations towards the $\Ie{0}$ curve as the extra energy is progressively radiated away by fast MHD waves. Such results have been reported in both 2D and 3D numerical simulations of prominence oscillations \citep[\textit{e.g.},][respectively]{Schutgens:1999,Zhou:2018}.

%  {\S}{\bf --- Additional solutions} \\
Furthermore, the $I_\mathrm{evol}$ curve can also cross the other branch of the $\Ie{0}$ curve past the point C$'$, such as at points A$''$ and B$''$ in Figure~\ref{fig:draining}.  
Since this part is unstable (see $f$ arrows in Figure~\ref{fig:draining}), there is the possibility of an eruption if the system has sufficient energy to reach this unstable part.  This region is indicated qualitatively with a pink area in Figure~\ref{fig:draining}. Its extension towards the side with small $h$ values is limited by the ability of the magnetic force to decrease the kinetic energy before the unstable region, at larger $h$ values, is reached.  We will not study this aspect any further since it is expected to be an effect localised to the family of $I_\mathrm{evol}$ curves near to point C$'$ and this would need a detailed analysis (it depends both on $m$ and $\Be (h)$).  We only point out that an eruption may be started, by draining mass, before the line current evolves to the limiting curve $I_\mathrm{evol}$ that passes the first unstable point, C$'$, of the $\Ie{0}$ curve.    

%\newpage
%------------------------------------------------------
\subsection{Modification of the Equilibrium Height}
    \label{SS:equ_height}

In this subsection, we give a quantitative estimate of the ideas described with Figure~\ref{fig:draining}.
Specifically, we analyse the mass-draining from the equilibrium located at $(\hm,\Im)$ with mass $m$, to the equilibrium located at $(\ho,\Io)$ without mass.

%  {\S}{\bf --- Flux conservation} \\
The total flux passing between the bottom of the flux rope and the surface is,
\BA
    F(h)=\frac{\mu_\mathrm{0} I}{2\pi}\,\ln\left(\frac{2h}{a}\right)-\Int{0}{h-a}B_\mathrm{ext}(z) \dif z\,. \label{eq_F_general}
\EA

Conserving flux passing below the flux rope per unit length $\Delta y$ during the mass-draining requires that $F(h_\mathrm{m})=F(h_\mathrm{0})$, hence,
  \BA   
 & & \Int{0}{\hm-a} \Be (z) \rmd z -\frac{\mu_\mathrm{0}}{2\pi}\, \Im \ln (2\hm /a) \nonumber \\
 &=& \Int{0}{\ho-a} \Be (z) \rmd z -\frac{\mu_\mathrm{0}}{2\pi}\, \Io \ln (2\ho /a) \label{eq_F_conserved} \, ,
  \EA
where we suppose that the flux rope radius, $a$, is small compared to its height, and that $a$ remains unchanged by the mass-draining to simplify the expressions as evolution in $a$ has a low effect on the results (similar to the case $m=0$ in DA10 where $a$ did not evolve). Equation~(\ref{eq_F_conserved}) explicitly states that the two equilibrium are on the same $I_\mathrm{evol}(h)$ curve (\eq{Ievol_general})

%  {\S}{\bf --- Cases of small mass} \\
We next suppose that the two equilibria $(\hm,\Im)$ and $(\ho,\Io)$ are close enough, so that the mass has a small effect on the force balance ($m\, g\, h\ll(\pi/\mu_\mathrm{0})(h \Be)^2$).  We also take the equilibrium without mass as a reference to express all terms of the Taylor development and define the variation quantities: $\Delta h = \ho-\hm$,  $\Delta I = \Io-\Im$.  From Figure~\ref{fig:draining}, $\Delta h >0$ and $\Delta I<0$.  

%  {\S}{\bf --- Flux conservation} \\
With a Taylor development to first order in $\Delta h$ and 
$\Delta I$ of \eq{F_conserved}, the conservation of flux imposes the relationship,
  \BE   \label{eq_F_conserved_Delta}
  \frac{\Delta h}{\ho}  = - \frac{\mu_\mathrm{0}}{2\pi}\, \ln (2\ho /a)\,  \frac{\Delta I}{\Io}  \, .
  \EE

%  {\S}{\bf --- Force balance change: method} \\
%The equilibrium evolution is analysed by a Taylor expansion in  $\Delta h, \Delta I$ and $m$ of \eq{Ie}. Then, inserting \eq{F_conserved_Delta} to eliminate $\Delta I$ provides $\Delta h$ as a function of $m$. However, it is more useful to apply such a Taylor %expansion to the more general \eq{f} as this provides the relation between the draining itself, and the test of stability
%of the equilibrium after the draining.

%  {\S}{\bf --- Force balance change} \\
The equilibrium curve without mass satisfies,
  \BE   \label{eq_equil0}
  0 = \frac{\mu_\mathrm{0}\Io^2}{4\pi\ho} -\Io \Be (\ho) \, ,
  \EE
and the force balance with mass $m$ satisfies,
  \BE   \label{eq_pertf}
  \Delta f = \frac{\mu_\mathrm{0}\Im^2}{4\pi\hm} -\Im \Be (\hm) -m\, g \, .
  \EE
With $\Delta I$ rewritten as a function of $\Delta h$ with the flux conserved, \eq{F_conserved_Delta}, with $\Io =(4\pi/\mu_\mathrm{0})\ho\, \Be(h_\mathrm{0})$, the first order expansion around $(\ho,\Io)$ of \eq{pertf} is,
  \BA  
   \Delta f 
   &=&  -m\, g  + \Delta h\,\frac{4\pi}{\mu_\mathrm{0}}\, \Be^2      \label{eq_pertf_linear} \\
   &&  \left(1 + \frac{2\pi}{\mu_\mathrm{0} \ln (2\ho /a)}  
            +\left. \frac{\partial \ln \Be (h)}{\partial \ln h}\right|_{h=\ho}
             \right)
    \, .  \nonumber
  \EA
  
  \begin{figure*}[ht!]
\centerline{\includegraphics[width=1.05\textwidth, clip=, trim=0 0 0 0]{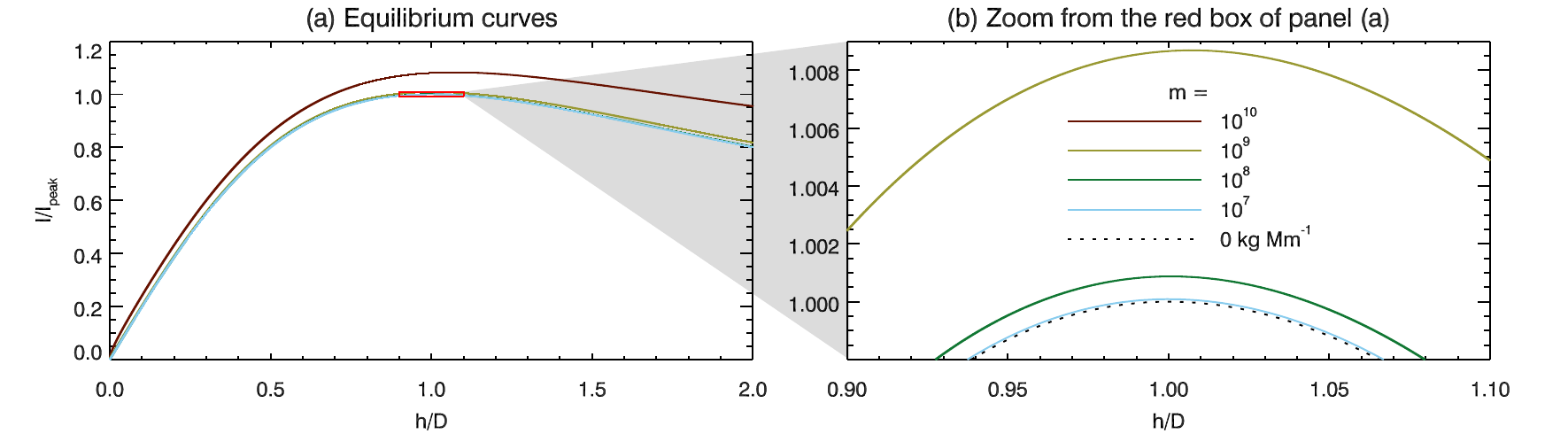}
			}
            \caption{Equilibrium curves demonstrating \eq{i_bip_m}, the relationship between electric current magnitude and height of the line current suspended within a bipolar background potential magnetic field generated by a 4~G mean surface field. These equilibrium curves are calculated assuming a range of prominence mass between $10^{7}$ -- $10^{10}$~kg~Mm$^{-1}$. The dotted-black line corresponds to no mass within the system, comparable to the solid-black line in Figure~2c of \citet{Demoulin:2010}.}\label{fig:instabilitywithwithoutmass_bipole}
\end{figure*}

%  {\S}{\bf --- Stability analysis} \\
With $m=0$, \eq{pertf_linear} describes the test of stability of the equilibrium around the point $(\ho,\Io)$. Next, we introduce the notations,
  \BE   \label{eq_n}
  n = \left. - \frac{\partial \ln \Be (h)}{\partial \ln h}\right|_{h=\ho} \, ,
  \EE
for the negative logarithmic derivative of the external field component, commonly referred to as the decay index \citep{Bateman:1978,Filippov:2001,Torok:2005,Zuccarello:2016}, and,
  \BE   \label{eq_ncrit}
  \ncrit = 1 + \frac{2\pi}{\mu_\mathrm{0} \ln (2\ho /a)} \, ,
  \EE
which is Eq. (33) of DA10 (with $n_{\rm a} =0$ since we have a fixed $a$ value).
Then, \eq{pertf_linear} is rewritten as,  
  \BE  \label{eq_pertf_n}
   \Delta f =  - m\, g  + \Delta h\,\frac{4\pi}{\mu_\mathrm{0}}\, \Be^2\,  (\ncrit -n)  \, .  
  \EE
With $m=0$, the equilibrium at $(\ho,\Io)$ is stable if $\Delta f$ is oppositely directed to the displacement $-\Delta h$ from $\ho$ to $\hm$. This is achieved for $n<\ncrit$, as expected. 

%  {\S}{\bf --- Draining mass} \\
Supposing that the extra energy is somehow dissipated, \textit{i.e.}, $\Delta f=0$, \eq{pertf_n} also describes the mass-draining from the equilibrium at $(\hm,\Im)$ to the equilibrium at $(\ho,\Io)$.  This draining implies the shift in height,
  \BE  \label{eq_dh_draining}
   \Delta h =  \frac{m\, g}{\frac{4\pi}{\mu_\mathrm{0}}\,\Be^2\,  (\ncrit -n)}   \, , 
  \EE
to the new equilibrium $(\ho,\Io)$ which exists only for $n<\ncrit$.
This quantifies the graphical description of Figure~\ref{fig:draining}. In particular, it shows that $\Delta h$ is proportional to the loaded mass $m$ and inversely proportional to distance, in terms of decay index, to the loss-of-equilibrium point ($n=n_\mathrm{crit}$).
Finally, the strength of the external field has a strong effect on $\Delta h$ since a factor 10 on $\Be$ decreases $\Delta h$ by a factor 100 (this factor 10 on $\Be$ is the order of magnitude for the ratio between the field present in active and quiescent prominences for example). We conclude that the draining of a given mass $m$ could cause the height of the prominence to increase from a tiny to a very large amount (up to the loss-of-equilibrium and resulting eruption) depending on precisely where this draining occurs along the equilibrium path and on the strength of the external field.

\section{Results}
	\label{S:Results}

\subsection{Bipolar Background Field}
	\label{SS:Bipole}

We begin by expanding on the case investigated in DA10 to explore the effect of including mass on the evolution of the line current, suspended within a bipolar background magnetic field, up to its loss-of-equilibrium. Here, the bipolar background magnetic field is supplied by two, infinitely long polarities at distance $\pm~D$ from the position of the line current (DA10),
\begin{equation}
	B_\mathrm{ext} = 2 \phi D (\pi(h^2+D^2))^{-1}\label{eq_b_bip} \, ,
\end{equation}
where $\phi$ is the magnetic flux per unit length in the invariant direction. Substituting Eqs.~(\ref{eq_b_bip}) and (\ref{eq_b_image}) into (\ref{eq_f}), we arrive at the condition for the system in equilibrium with $f=0$,
\begin{equation}
	\frac{\mu_\mathrm{0}I^2}{4\pi h} - \frac{2 \phi D I}{\pi (h^2 +D^2)}-mg=0 \label{eq_bip} \, .
\end{equation}
The equilibrium curve for the massless line current is,
\begin{equation}
	\frac{I_\mathrm{eq,0}(h)}{I_\mathrm{peak}}=\tilde{I}_\mathrm{eq,0}(\tilde{h})= \frac{2\tilde{h}}{(\tilde{h}^2+1)} \label{eq_i_bip} \, ,
\end{equation}
where $I_\mathrm{eq,0}(h)$ is normalised by its maximum value, $I_\mathrm{peak}=\frac{\phi}{\pi}$, occurring at height $\tilde{h}_\mathrm{peak}=h_\mathrm{peak}/D=1$. Note that Equation~(\ref{eq_i_bip}) corrects a typo of DA10. For the case where a line current does contain mass, $\tilde{I}_\mathrm{eq,m}(\tilde{h})$ takes the form similar to Eq.~(\ref{eq_Ie}),
\begin{equation}
	\tilde{I}_\mathrm{eq,m}(\tilde{h})= \frac{2\pi^2\, \tilde{h} D}{\mu_\mathrm{0}\phi} \Biggl(B_\mathrm{ext} + \sqrt[]{\displaystyle \left(B_\mathrm{ext} \right)^2 +\frac{\mu_\mathrm{0} mg}{\pi \tilde{h}D}}\Biggr)\label{eq_i_bip_m} \, .
\end{equation}
In Figure~\ref{fig:instabilitywithwithoutmass_bipole} we show a comparison between normalised equilibrium curves of line currents suspended within a ``typical'' quiet-Sun region of average surface field strength equal to 4~G and loaded with a range of masses. The properties of the masses used are presented in Table~\ref{table:mass_preperties}, assuming a typical quiescent prominence of dimensions: length~$=$~100~Mm, height~$=$~30~Mm, width~$=$~4~Mm \citep{Labrosse:2010,Xia:2012}.

%\newpage
\subsection{Effect of Mass on Line Current Equilibrium}
	\label{SS:LCEquilibrium}

\begin{figure*}[ht!]
\centerline{\includegraphics[width=1.05\textwidth, clip=, trim=0 0 0 0]{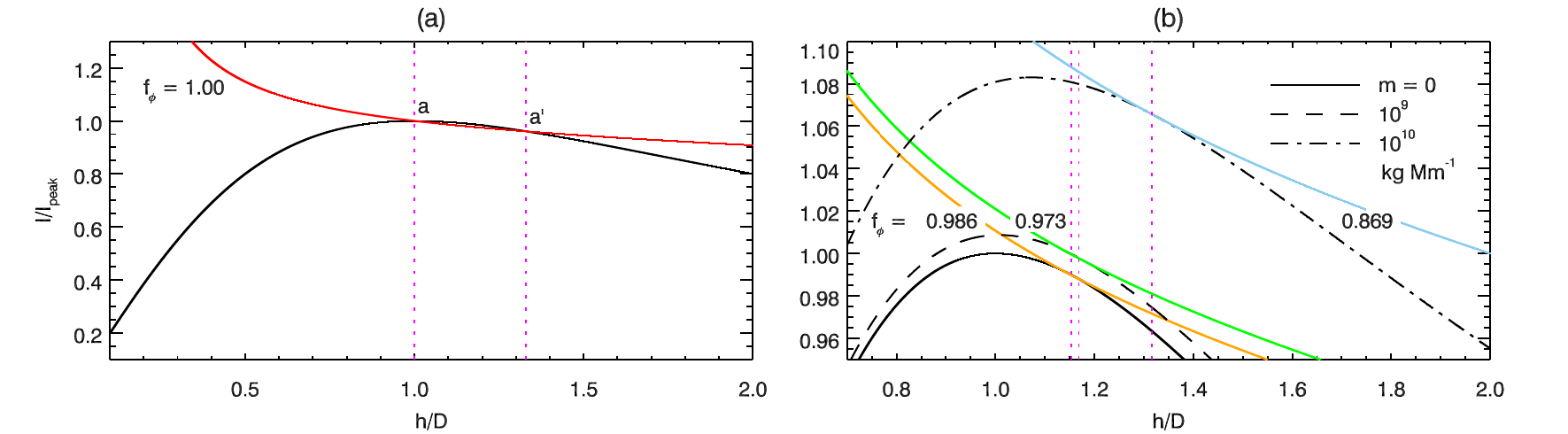}
			}
            \caption{The effect of mass on the stability of a line current suspended within a bipolar background potential field. \textit{Panel a}; intersection of $\tilde{I}_\mathrm{eq,0}(\tilde{h})$ and $\tilde{I}_\mathrm{evol}(\tilde{h})$ with $f_\phi=1$ indicating two equilibrium positions, $\tilde{h}=1,1.33$, as in Figure~2c of \citet{Demoulin:2010}. \textit{Panel b}; orange, green, and blue curves correspond to the last point-of-intersect, with $f_\phi$ decreasing, between $\tilde{I}_\mathrm{eq,m}(\tilde{h})$ and $\tilde{I}_\mathrm{evol}(\tilde{h})$, with $f_\phi=0.986,0.973$ and $0.869$ for line currents loaded with mass equal to 0, $10^{9}$, and $10^{10}$~kg~Mm$^{-1}$, respectively. The vertical magenta-dotted lines indicate the $h/D$ value for this last intersection, in each case, between the two $\tilde{I}(\tilde{h})$ curves. $h/D$ value is seen to increase as more mass is loaded.}\label{fig:ftheta}
\end{figure*}

Here, we impose the same flux evolution analysis, described in Section~\ref{S:Mass_draining}, on the equilibrium curves presented in Figure~\ref{fig:instabilitywithwithoutmass_bipole} to study the effect of mass on the equilibrium of the host line current. 

\begin{table}
\vspace{0.03\textheight}
\begin{center}
%\resizebox{0.3\textwidth}{!}{%
\begin{tabular}{ccc}
\hline
      \multicolumn{1}{c}{$N_\mathrm{H}$ (Total)} &
      \multicolumn{1}{c}{Mass (Total)} &
      \multicolumn{1}{c}{Mass (Per unit length)}\\
    (cm$^{-3}$)                     &     (kg)      &    (kg~Mm$^{-1}$)     \\
\hline
    5~$\times$~10$^7$               &   10$^{9}$    &   10$^{7}$           \\
    5~$\times$~10$^8$               &   10$^{10}$   &   10$^{8}$           \\
    5~$\times$~10$^9$               &   10$^{11}$   &   10$^{9}$           \\
    5~$\times$~10$^{10}$            &   10$^{12}$   &   10$^{10}$          \\

\hline
\end{tabular}%}
\end{center}
\caption{The properties of the masses loaded onto the line currents presented in Figure~\ref{fig:instabilitywithwithoutmass_bipole}. It is assumed that all mass within a prominence is cool (low ionisation ratio), therefore $N_\mathrm{H}$ is the number density of neutral hydrogen assuming the range of masses within the second column \citep{Labrosse:2010}.} \label{table:mass_preperties}
\end{table}     

The reference state with fluxes $\tilde{F}_0$ and $\phi_0$ is defined at the maximum of the $\tilde{I}_\mathrm{eq,0}(\tilde{h})$ curve (DA10, and references therein),
\begin{equation}
	\displaystyle \tilde{F}_0=\frac{\mu_\mathrm{0}}{2\pi}\tilde{I}_\mathrm{peak} \,\mathrm{ln}\left(\frac{2\tilde{h}_\mathrm{peak}}{\tilde{a}}\right)-2\mathrm{tan}^{-1}\left(\tilde{h}_\mathrm{peak}-\tilde{a}\right) \, , \label{eq_f_norm}
\end{equation}
where $\tilde{I}_\mathrm{peak}$ is the maximum value of $\tilde{I}_\mathrm{eq,0}(\tilde{h})$, and $\tilde{a}$ is the normalised radius of the line current ($\tilde{a}=a/D = 0.1$ hereafter). We readily find $\tilde{I}_\mathrm{evol}(\tilde{h})$ from \eq{Ievol_general},
\begin{equation}
	\displaystyle \tilde{I}_\mathrm{evol}(\tilde{h})=\frac{\tilde{F}+2\mathrm{tan}^{-1} \left(\tilde{h}-\tilde{a}\right)}{\frac{\mu_\mathrm{0}}{2\pi}\,\mathrm{ln}\left(\frac{2\tilde{h}}{\tilde{a}}\right)} \, ,
\end{equation}
where $\tilde{F}=\tilde{F}_0/f_\phi$. 

The intersection of $\tilde{I}_\mathrm{evol}(\tilde{h})$ and $\tilde{I}_\mathrm{eq,0}(\tilde{h})$ for the case of $f_\phi=1$ for the massless line current is shown in Figure~\ref{fig:ftheta}a. The orange curve in Figure~\ref{fig:ftheta}b then corresponds to $f_\phi=0.986$ (1.4$\%$ reduction in $\phi_0$, the strength of the photospheric polarities) applied, also, to the case of a massless line current, indicating a single point of intersection between the two $\tilde{I}(\tilde{h})$ curves, at $h/D = 1.15$. Any further reduction in $f_\phi$ results in no intersection between the two $\tilde{I}(\tilde{h})$ curves. DA10 demonstrate that such a line current experiences an ideal-MHD instability and an outward force drives the eruption of the line current.

In Figures~\ref{fig:instabilitywithwithoutmass_bipole} and \ref{fig:ftheta}b it is shown that an increase in the amount of mass loaded onto the line current results in a shift in the maximum value of $I/I_\mathrm{peak}$ and its corresponding $h/D$ value. As with the orange curve, the green and blue curves are the last point of intersect between $\tilde{I}_\mathrm{eq,m}(\tilde{h})$ and $\tilde{I}_\mathrm{evol}(\tilde{h})$ where a line current is loaded with $10^{9}$ and $10^{10}$~kg~Mm$^{-1}$, respectively. This implies that the flux of the photospheric polarity must decrease further than for the massless case in order for the mass-loaded line current to experience an ideal-MHD instability. For a line current loaded with $10^{9}$ or $10^{10}$~kg~Mm$^{-1}$, ideal-MHD instability occurs after $\phi$ has decreased by $2.7\%$ and $13.1\%$, respectively, at a height of $h/D=1.17,1.32$. Therefore, the simple model presented here appears to demonstrate that a mass-loaded line current can be significantly anchored as a result of the inclusion of mass \citep[cf.][]{Blokland:2011}, requiring additional current within, photospheric flux decay below, and height for the line current to experience loss-of-equilibrium.

\citet{Fan:2018a} recently published the first example in which a prominence comprised of mass on the order of $10^{12}$~kg erupted in a fully MHD simulation. Interestingly, the existence of this prominence was shown to have a significantly stabilising effect on its host flux rope when compared to an identical flux rope without prominence formation induced. Specifically, the prominence was shown to inhibit the initiation of the kink instability prior to a successful eruption. The work presented here shows that a similar conclusion can also be reached with the torus instability using significantly simplified conditions.

%\newpage
\subsection{Effect of Mass-Draining on the Pre-Eruptive Evolution of the Line Current}
    \label{SS:mass_draining}

\citet{Blokland:2011} showed that the inclusion of mass within their MHS model caused the center of their flux rope to be pulled downwards \textit{i.e.}, the Shafranov shift. Further to this, we have established that the inclusion of mass within the simple model presented by \citet{vanTend:1978} and expanded by DA10, can result in a non-negligible modification to the equilibrium curves and implies additional stability. It is therefore reasonable to suggest that the removal of this mass from a pre-loss-of-equilibrium line current will also result in a modification to its evolution, as suggested in several observational case studies of prominences \citep[\textit{e.g.},][]{Seaton:2011,Bi:2014,Reva:2017,Jenkins:2018}.

To test this hypothesis and simulate the draining of prominence mass from a flux rope, we first apply the general, first-order development described in Section~\ref{S:Mass_draining}, specifically \eq{dh_draining}, to the specified bipolar background magnetic field of Eq.~(\ref{eq_b_bip}). The results are presented in Figure~\ref{fig:dh_bipole_obs} as the dashed-black lines. $\Delta h$ is larger when $h_\mathrm{m}$ is closer to the loss-of-equilibrium point (\textit{i.e.}, $n=n_\mathrm{crit}$). However, \eq{dh_draining} is derived with a Taylor expansion in $\Delta h$, so it cannot describe large $\Delta h$ values.

%Here we can see that the draining of mass will increasingly influence the line current up until the height where there is no intersection between $\tilde{I}_\mathrm{evol}(\tilde{h})$ and $\tilde{I}_\mathrm{eq,0}(\tilde{h})$; the more mass drained from a line current, the larger the resulting change in height. At the point of no intersection between $\tilde{I}_\mathrm{evol}(\tilde{h})$ and $\tilde{I}_\mathrm{eq,0}(\tilde{h})$, it is believed that the line current would experience loss-of-equilibrium; the mass-draining will have triggered the eruption of the line current. However, manual inspection of Figure~\ref{fig:ftheta}b details that the orange $\tilde{I}_\mathrm{evol}(\tilde{h})$ curve ($f_\phi=0.986$) intersects the $\tilde{I}_\mathrm{eq,m}(\tilde{h})$ curve for a line current loaded with $m=10^5$~g~cm$^{-1}$ at $h/D\approx0.8$. However, the last solution to \eq{dh_draining}, according to Figure~\ref{fig:dh_bipole_obs}, is at $h/D\approx0.9$. In addition, from the same equation we expect the $\Delta h$ value to tend towards infinity as the line current approaches its loss-of-equilibrium ($n=n_\mathrm{crit}$). This discrepancy suggests that the analytical solution is unable to accurately describe the evolution of the line current when both considering large prominence masses, and as the line current approaches loss-of-equilibrium.

Therefore, we have used the ``Chebfun'' package \citep[see,][]{Driscoll:2014} implemented in MATLAB to solve numerically for the intersects between $\tilde{I}_\mathrm{eq,m}(\tilde{h})$, $\tilde{I}_\mathrm{eq,0}(\tilde{h})$, and $\tilde{I}_\mathrm{evol}(\tilde{h})$ for a range of values of $f_\phi$. These solutions are presented in Figure~\ref{fig:dh_bipole_obs}, plotted over the analytical solution for comparison. Although the main trend is accessible via both the analytical and numerical solutions, the numerical solution emphasises the sensitivity of the equilibrium to mass evolution when the line current is close to its loss-of-equilibrium.

\section{Implications for Observations}
	\label{S:Implications}

\begin{figure}
\centerline{\includegraphics[width=0.5\textwidth,clip=, trim=20 0 0 0]{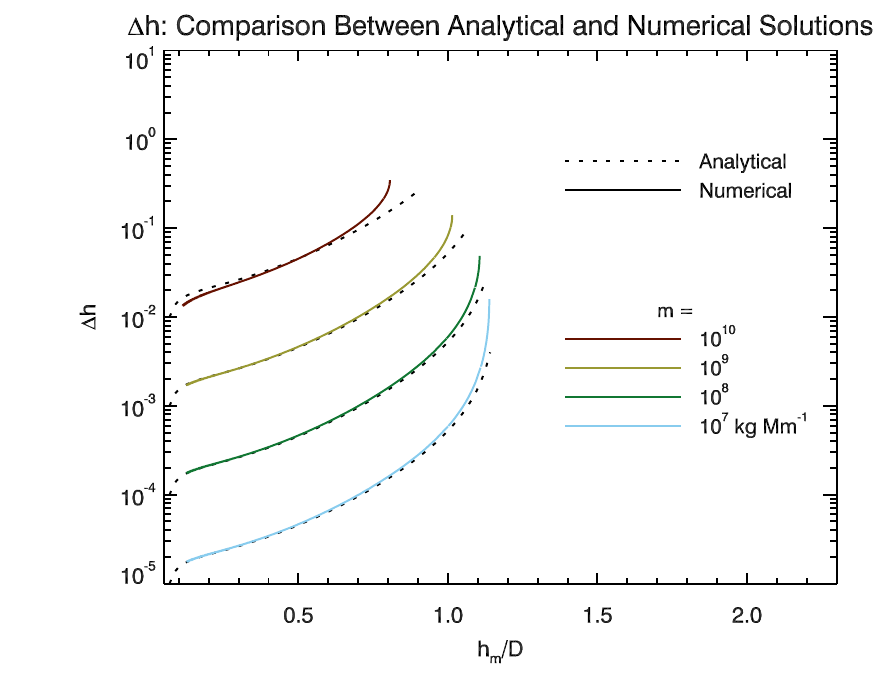}}
\caption{The change in the height of a line current due to a range of mass-draining, assuming a bipolar background potential magnetic field generated by an average surface field of strength 4~G. Analytical solutions to \eq{dh_draining} are plotted for each mass as dashed-black lines. Overplotted on these dashed lines are the solid-coloured lines representing the numerical solution. The analytical solution works well for small $h$ and $m$ values, but clearly deviates from the numerical solution at larger values.} \label{fig:dh_bipole_obs}
\end{figure}

We now move to establish a basic comparison between the results of the above model and two specific observations of mass-draining. Our model shows that some of the quantities may be very sensitive to the value used in their computation, see \textit{e.g.}, Figure~\ref{fig:dh_bipole_obs} for large $h_\mathrm{m}/D$. The model input parameters (filament dimensions, height, mass, and external field as a function of time) require indirect, often complex methods to be estimated from observations, and are subject to different types of errors. Therefore, our intention is to establish an order of magnitude indication to the importance of mass-draining in these two cases, not an exact measure. Furthermore, we find that varying the value of $a/D$ between 0.1 and 0.5 results in modifications to the stability of the line current of only a few \%. Therefore, for this comparison we maintain the assumption of a the thin flux rope and fix $a/D=0.1$.

%The errors on any observationally-constrained inputs to the model described here are understandably large. Therefore,

\begin{figure}
\centerline{\includegraphics[width=0.5\textwidth,clip=, trim=20 0 0 0]{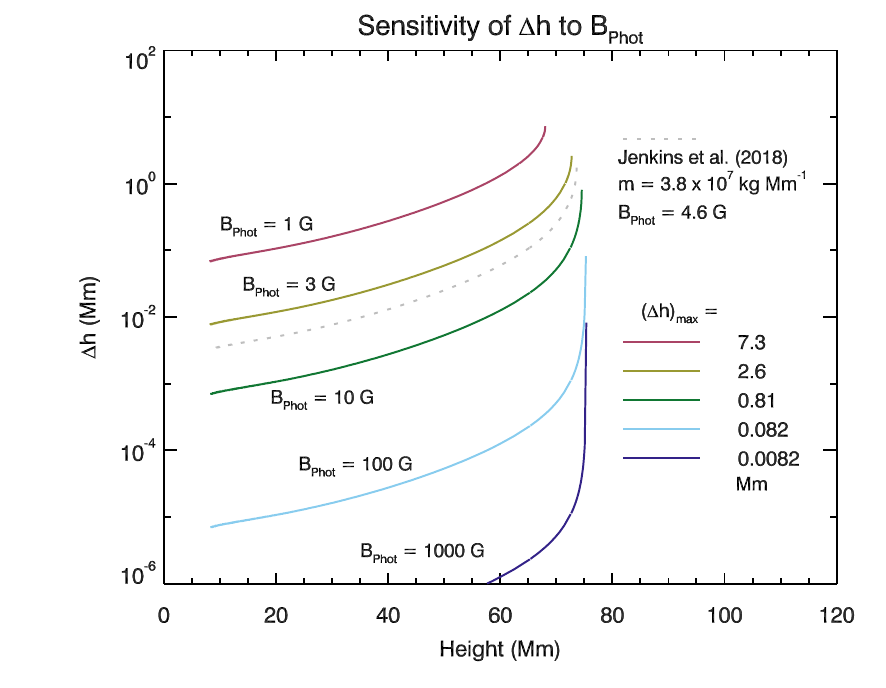}
            }
\caption{The modification to the height of the line current assuming a draining equal to  1~$\times$~10$^{10}$~kg of prominence mass \citep{Jenkins:2018} at a range of photospheric magnetic field strengths. Quiet-Sun surface field strengths result in a significantly larger change in height due to mass-draining than field strengths similar to those observed in active regions.}\label{fig:phi_sens}
\end{figure}

We first refer to a recent case study by \citet{Jenkins:2018} in which the authors used the column density estimation technique of \citet{Williams:2013} and \citet{Carlyle:2014a} to study the draining of mass from an erupting quiet-Sun prominence. According to the authors observations, shortly prior to the prominence's eruption the total mass within the field-of-view reduced by at least 1~$\times$~10$^{10}$~kg, equal to 15\% of the initial mass within the field-of-view. The additional properties of the erupting prominence were estimated from observations taken using the Atmospheric Imaging Assembly \citep[AIA;][]{Lemen:2012} on board the Solar Dynamics Observatory \citep[SDO;][]{Pesnell:2012},
\begin{align*}
   			    D   &= 65~\mathrm{Mm}\\
				L_\mathrm{y}	&= 260~\mathrm{Mm}\\
    B_\mathrm{Phot} &= 4.6~\mathrm{G} \\
    		\phi	&=   \frac{\pi D B_\mathrm{Phot}}{2}\\
         \delta m	&= 1~\times~10^{10}~\mathrm{kg}
\end{align*}
where $L_\mathrm{y}$ is the length of the prominence. Specifically, $D$ is half the width, and $L_\mathrm{y}$ the length, of the red-dashed box in Figure~3b of \citet{Jenkins:2018}. The value of $B_\mathrm{Phot}$ is the average strength of the magnetic field within the bounds of the red-dashed box in Figure~3a of \citet{Jenkins:2018}. The results of the application of these values to the model are shown in Figure~\ref{fig:phi_sens}. 

In the application of the observations to this model we have set initial mass equal to 9~$\times$~10$^{10}$~kg and final mass equal to 8~$\times$~10$^{10}$~kg. According to the model, such a mass-loaded line current would need to reach a height of $\approx$~75~Mm to lose stability, $\approx$~30~Mm higher than the prominence top was observed; the quiet-Sun prominence was suggested to lose equilibrium, inferred by the large acceleration, after it had risen to a height of $\approx$~45~Mm. In fact the comparison between model and observations cannot be precise because of the approximate values derived from observations and the simplicity of the model. Moreover, all of the mass present in the model exists at the height of the line current, a location representative of the axis of a flux rope. As it is commonly assumed that prominence material resides below this height, in the dips of the magnetic field of a flux rope \citep[\textit{e.g.},][]{Aulanier:1998b,Gunar:2015}, we expect the model height corresponding to loss-of-equilibrium to always be larger than any observed prominence height \citep[see also,][]{Zuccarello:2016}.

The increase in height observed by \citet{Jenkins:2018} after the prominence underwent mass-draining was $>$~60~Mm before leaving the field-of-view. The simple model described here predicts the maximum possible increase in height for the same amount of mass-draining to be up to 1.7~Mm, assuming the final state is also in equilibrium. However, it is suggested by the authors that the flux rope associated with the prominence was at a point of marginal instability when the mass-draining initiated. Indeed, we have shown that the simple model described here predicts the largest increases in height due to mass-draining to occur as the line current approaches its loss-of-equilibrium. Hence, the large increase in the height of the observed prominence (60 Mm) shortly after the draining of mass could be interpreted as being caused by the flux rope losing equilibrium and erupting into the heliosphere due to the torus instability. The prominence observed by \citet{Jenkins:2018} did successfully erupt and was later observed as a CME by multiple coronagraphs.

The mass estimates of the prominence material studied by \citet{Jenkins:2018} were derived from observations captured using the Extreme Ultraviolet Imager \citep[EUVI;][]{Wuelser:2004} on board the Solar Terrestrial Observatory Behind \citep[STEREO;][]{Kaiser:2008} spacecraft. At the time of the observations, December 2011, EUVI was capturing high temporal resolution images in only the 195~\AA\ passband; the other filters were at a much lower cadence. For this reason, the column density of the prominence was calculated using the so-called `monochromatic method', resulting in a lower-limit estimate to the column density. Therefore, we take the derived value of total mass and mass-drained as lower-limits, and in turn all values of $\Delta h$ to be lower-limit estimates to the increase in the height of the line current.

Next, we compare to an earlier case study, presented by \citet{Seaton:2011}, in which it was concluded that mass-draining from a prominence rooted within an active region was responsible for the $\approx$~35~Mm height rise prior to the eruption of the prominence. The active region that the eruptive prominence was located in was in its decaying phase, with an average surface magnetic field strength of $\gtrsim$~100~G according to magnetogram observations taken using the Michelson Doppler Imager \citep[MDI;][]{Scherrer:1995} on board the Solar and Heliospheric Observatory \citep[SOHO;][]{Domingo:1995}. Our model can be used to test this conclusion by assuming the same degree of mass-draining as was observed by \citet{Jenkins:2018}, and modifying $B_\mathrm{Phot}$ so as to test the sensitivity of the model to a range of surface fluxes. 

In the solar context, higher values of $B_\mathrm{Phot}$ are associated with smaller values of $D$, in turn reducing the critical height of the flux rope. Indeed, this is a commonly observed and well studied relationship between prominence height and magnetic domain \citep[\textit{e.g.},][and references therein]{Rompolt:1990,McCauley:2015,Filippov:2016}. However, in order to meaningfully vary $D$ with $B_\mathrm{Phot}$ within this model, additional assumptions would have to be made. Therefore, to facilitate a simple comparison between the two observational case-studies and the additional range of realistic surface flux values, we opt to compare conditions for a ``normalised filament'', fixing $D$ as in \citet{Jenkins:2018} and simply varying $B_\mathrm{Phot}$.

The results, shown in Figure~\ref{fig:phi_sens}, suggest that increasing the surface magnetic flux results in a stronger background potential magnetic field, and reduces the effect that mass-draining can have on the height change $\Delta h$. According to the model, draining 1~$\times$~10$^{10}$~kg from a line current embedded within a bipolar background potential magnetic field that has a surface magnetic field strength of $\gtrsim$~100~G would result in a very small maximum change to the height of the line current unless the configuration is very close to loss-of-equilibrium. It is therefore unlikely that the mass-draining was directly responsible for the observed $\approx$~35~Mm increase in the height of the prominence. Nevertheless, as appears to have been the case in the prominence studied by \citet{Jenkins:2018}, the mass-draining may have been responsible for upsetting the equilibrium towards the non-equilibrium point.

A similar result to this has previously been reported by \citet{Reeves:2005}, in which the authors concluded the effect of mass was likely to be negligible in a system restricted by a background field stronger than 6~G. Our result is complementary to this by providing a quantitative comparison for a range of surface fluxes and masses.

%\newpage
\section{Discussion and Summary}
	\label{S:Conclusions}
	
%	\section{Discussion}
%    \label{S:Discussion}
The general cases described in this manuscript detail how the inclusion of realistic prominence masses and complete draining of this mass from the line current can have both stabilising and destabilising effects. Returning to Figure~\ref{fig:draining} for comparison, a line current at point A, for example, will drain total mass and move along the constant $F$ curve to A$'$, resulting in damped oscillations around A’. In such a case, the line current does not experience a loss-of-equilibrium; the draining of mass has simply allowed the line current to increase in height to a new equilibrium and further evolution of other parameters would be required for a successful eruption to occur. The mass draining can also be partial and can also occur during the oscillations. Indeed, \citet{Zhou:2018} showed that the oscillation of the prominence in their 3D MHD simulation resulted in the draining of mass from the structure due to the periodic increase in height of field lines during the oscillation. The authors also note that this causes the height of individual field lines to increase due to the reduction in the gravitational force, although this is studied locally for a few field lines.

Considering, now, a line current evolving from C to C$'$ due to mass-draining, the line current would become unstable to an ideal-MHD instability as it reaches point C$'$, and experience a loss-of-equilibrium triggered by the draining of mass. 

For a line current that drains a partial amount of the total mass loaded, the height of the line current will increase accordingly, as has already been discussed in Section~\ref{SS:mass_draining}. If this is realised at point A or B of Figure~\ref{fig:draining}, the line current will not evolve all the way to point A$'$ or B$'$, rather a point on the constant $F$ curve that is in-between and dependent on the degree of draining. 

Considering point D, a point that is not sampled using the methods outlined in this manuscript, then the partial draining of total mass may result in the line current either reaching a stable equilibrium again or experiencing a loss-of-equilibrium. In this case, the nature of the line current post mass-draining would depend on the degree of mass drained. Graphically, for a line current to experience loss-of-equilibrium the constant $F$ curve cutting the mass-loaded line current equilibrium curve at point D would have to touch the mass-drained equilibrium curve tangentially or not at all. If we define $m_\mathrm{drained}$ as the amount of mass drained and $m_\mathrm{min}$ as the minimum amount of mass-draining required to destabilise a line current at point D, then if $m_\mathrm{drained}<m_\mathrm{min}$ the final state of the line current would be in equilibrium. Assuming no more mass-draining occurred, the additional physical parameters of the system would be required to evolve for a successful eruption to occur. It then follows that if $m_\mathrm{drained} \geq m_\mathrm{min}$ the line current would experience loss-of-equilibrium as a result of the mass-draining.

At point E the line current is already unstable to an ideal-MHD instability without any draining of mass. If mass draining was to occur at this point, the draining of total or partial mass would not contribute to the initiation of the loss-of-equilibrium but would instead contribute an additional accelerating force to the erupting flux rope.
	
Finally, applying specific conditions to the general case, it is shown that:
\begin{itemize}
    \item For a line current suspended within a bipolar background field generated by a surface field of 4~G, the inclusion of typical prominence masses can increase the height that the line current experiences an ideal-MHD instability by up to 14\%, indicating that the mass provides a larger anchoring effect than is typically assumed.
    \item The draining of the larger masses from a line current can cause a non-negligible increase in the height of the line current without upper bound, with the largest height increase observed as the line current approaches its loss-of-equilibrium.
    \item Using the observational measurements of \citet{Jenkins:2018} as the input parameters, it is shown that the modification to the height of the line current due to mass-draining is as much as 1.7~Mm. This non-negligible increase in the height of the line current effectively demonstrates the ability for mass-draining to perturb the equilibrium of weak field quiescent flux ropes.
    \item Scaling the model for comparison with observations presented by \citet{Seaton:2011}, it is shown that draining mass from a line current suspended in a background field generated by up to kilogauss surface field results in only a negligible modification to the height of the line current.
\end{itemize}
We have discussed the role that mass plays in the \textit{global} evolution and eruption of flux ropes, suggesting that it depends on four main parameters; the strength of the surface field generating the background potential field, how much mass is loaded into a flux rope, how much mass drains during its evolution, and when along a flux rope's equilibrium curve the mass drains. The effect of the \textit{local} evolution of plasma within prominences is not discussed in this manuscript, \textit{i.e.}, the mass-draining that is studied here differs from the mass-loss due to the Rayleigh-Taylor instability (RTI) that has been studied extensively in both observations and simulations \citep[\textit{e.g.},][]{Hillier:2012,Xia:2016,Hillier:2018}. In addition, \citet{Kaneko:2018} pointed out that, in their case, the mass-loss from the prominence due to RTI was balanced by new condensations into the prominence. A parametric study would be required in order to ascertain the effect of such local evolutions of mass on the global stability of a flux rope--prominence system.

Finally, we conclude that the role of mass within solar eruptions, particularly those involving quiescent prominences, is greater than has been historically attributed, and requires a more in-depth analysis.

%\section*{}

\acknowledgments
The authors would like to thank the anonymous referee for their useful comments that improved the clarity of the article. We would also like to thank the members of the "Solving The Prominence Paradox" ISSI team for their discussions on this topic during their second meeting. JMJ thanks the STFC for support via funding given in his PhD Studentship, and travel funds awarded by the Royal Astronomical Society. MH thanks the University of Birmingham School of Mathematics for funding given in his PhD Studentship, and the University of Adelaide for funding given in his Beacon of Enlightenment Scholarship. GV acknowledges the support of the Leverhulme Trust Research Project Grant 2014-051. GA and PD thank the Programme National Soleil Terre of the CNRS/INSU for financial support. DML received support from the European Commissions H2020 Programme under the following grant agreements; GREST (no. 653982) and Pre-EST (no. 739500) and from the Leverhulme Trust as an Early-Career Fellow (ECF-2014-792). LvDG acknowledges funding under STFC consolidated grant number ST/N000722/1, Leverhulme Trust Research Project Grant 2014-051, and the Hungarian Research grant OTKA K-109276.

%\begin{multicols}{2}

\bibliography{ms}

\begin{thebibliography}{}
\expandafter\ifx\csname natexlab\endcsname\relax\def\natexlab#1{#1}\fi
\providecommand{\url}[1]{\href{#1}{#1}}
\providecommand{\dodoi}[1]{doi:~\href{http://doi.org/#1}{\nolinkurl{#1}}}
\providecommand{\doeprint}[1]{\href{http://ascl.net/#1}{\nolinkurl{http://ascl.net/#1}}}
\providecommand{\doarXiv}[1]{\href{https://arxiv.org/abs/#1}{\nolinkurl{https://arxiv.org/abs/#1}}}

\bibitem[{{Aulanier} {et~al.}(1998){Aulanier}, {D\'emoulin}, {van
  Driel-Gesztelyi}, {Mein}, \& {Deforest}}]{Aulanier:1998b}
{Aulanier}, G., {D\'emoulin}, P., {van Driel-Gesztelyi}, L., {Mein}, P., \&
  {Deforest}, C. 1998, \aap, 335, 309

\bibitem[{{Bateman}(1978)}]{Bateman:1978}
{Bateman}, G. 1978, {MHD instabilities} (Cambridge, Mass., MIT Press, 1978.~270
  p.)

\bibitem[{{B{\c a}k-St{\c e}{\'s}licka} {et~al.}(2013){B{\c a}k-St{\c
  e}{\'s}licka}, {Gibson}, {Fan}, {Bethge}, {Forland}, \&
  {Rachmeler}}]{Bak-Steslicka:2013}
{B{\c a}k-St{\c e}{\'s}licka}, U., {Gibson}, S.~E., {Fan}, Y., {et~al.} 2013,
  \apjl, 770, L28, \dodoi{10.1088/2041-8205/770/2/L28}

\bibitem[{{Bi} {et~al.}(2014){Bi}, {Jiang}, {Yang}, {Hong}, {Li}, {Yang}, \&
  {Yang}}]{Bi:2014}
{Bi}, Y., {Jiang}, Y., {Yang}, J., {et~al.} 2014, \apj, 790, 100,
  \dodoi{10.1088/0004-637X/790/2/100}

\bibitem[{{Blokland} \& {Keppens}(2011)}]{Blokland:2011}
{Blokland}, J.~W.~S., \& {Keppens}, R. 2011, \aap, 532, A93,
  \dodoi{10.1051/0004-6361/201117013}

\bibitem[{{Burlaga}(1988)}]{Burlaga:1988}
{Burlaga}, L.~F. 1988, \jgr, 93, 7217, \dodoi{10.1029/JA093iA07p07217}

\bibitem[{{Carlyle} {et~al.}(2014){Carlyle}, {Williams}, {van Driel-Gesztelyi},
  {Innes}, {Hillier}, \& {Matthews}}]{Carlyle:2014a}
{Carlyle}, J., {Williams}, D.~R., {van Driel-Gesztelyi}, L., {et~al.} 2014,
  \apj, 782, 87, \dodoi{10.1088/0004-637X/782/2/87}

\bibitem[{{Cheng} {et~al.}(2017){Cheng}, {Guo}, \& {Ding}}]{Cheng:2017}
{Cheng}, X., {Guo}, Y., \& {Ding}, M. 2017, Science in China Earth Sciences,
  60, 1383, \dodoi{10.1007/s11430-017-9074-6}

\bibitem[{{D{\'e}moulin}(1998)}]{Demoulin:1998}
{D{\'e}moulin}, P. 1998, in Astronomical Society of the Pacific Conference
  Series, Vol. 150, IAU Colloq. 167: New Perspectives on Solar Prominences, ed.
  D.~F. {Webb}, B.~{Schmieder}, \& D.~M. {Rust}, 78

\bibitem[{{D{\'e}moulin} \& {Aulanier}(2010)}]{Demoulin:2010}
{D{\'e}moulin}, P., \& {Aulanier}, G. 2010, \apj, 718, 1388,
  \dodoi{10.1088/0004-637X/718/2/1388}

\bibitem[{{D{\'e}moulin} {et~al.}(1991){D{\'e}moulin}, {Ferreira}, \&
  {Priest}}]{Demoulin:1991}
{D{\'e}moulin}, P., {Ferreira}, J., \& {Priest}, E.~R. 1991, \aap, 245, 289

\bibitem[{{D{\'e}moulin} \& {Priest}(1988)}]{Demoulin:1988}
{D{\'e}moulin}, P., \& {Priest}, E.~R. 1988, \aap, 206, 336

\bibitem[{{Dere} {et~al.}(1997){Dere}, {Brueckner}, {Howard}, {Koomen},
  {Korendyke}, {Kreplin}, {Michels}, {Moses}, {Moulton}, {Socker}, {St.~Cyr},
  {Delaboudini{\`e}re}, {Artzner}, {Brunaud}, {Gabriel}, {Hochedez}, {Millier},
  {Song}, {Chauvineau}, {Marioge}, {Defise}, {Jamar}, {Rochus}, {Catura},
  {Lemen}, {Gurman}, {Neupert}, {Clette}, {Cugnon}, {van Dessel}, {Lamy},
  {Llebaria}, {Schwenn}, \& {Simnett}}]{Dere:1997}
{Dere}, K.~P., {Brueckner}, G.~E., {Howard}, R.~A., {et~al.} 1997, \solphys,
  175, 601, \dodoi{10.1023/A:1004907307376}

\bibitem[{{Domingo} {et~al.}(1995){Domingo}, {Fleck}, \&
  {Poland}}]{Domingo:1995}
{Domingo}, V., {Fleck}, B., \& {Poland}, A.~I. 1995, \solphys, 162, 1,
  \dodoi{10.1007/BF00733425}

\bibitem[{Driscoll {et~al.}(2014)Driscoll, Hale, \& Trefethen}]{Driscoll:2014}
Driscoll, T.~A., Hale, N., \& Trefethen, L.~N. 2014, Chebfun Guide (Pafnuty
  Publications).
\newblock \url{http://www.chebfun.org/docs/guide/}

\bibitem[{{Fan}(2017)}]{Fan:2017}
{Fan}, Y. 2017, \apj, 844, 26, \dodoi{10.3847/1538-4357/aa7a56}

\bibitem[{{Fan}(2018)}]{Fan:2018a}
---. 2018, \apj, 862, 54, \dodoi{10.3847/1538-4357/aaccee}

\bibitem[{{Fan} {et~al.}(2018){Fan}, {Gibson}, \& {Tomczyk}}]{Fan:2018b}
{Fan}, Y., {Gibson}, S., \& {Tomczyk}, S. 2018, ArXiv e-prints,
  arXiv:1808.06142.
\newblock \doarXiv{1808.06142}

\bibitem[{{Fan} \& {Gibson}(2007)}]{Fan:2007}
{Fan}, Y., \& {Gibson}, S.~E. 2007, \apj, 668, 1232, \dodoi{10.1086/521335}

\bibitem[{{Filippov}(2018)}]{Filippov:2018}
{Filippov}, B. 2018, \mnras, 475, 1646, \dodoi{10.1093/mnras/stx3277}

\bibitem[{{Filippov}(2016)}]{Filippov:2016}
{Filippov}, B.~P. 2016, Geomagnetism and Aeronomy, 56, 1,
  \dodoi{10.1134/S0016793216010059}

\bibitem[{{Filippov} \& {Den}(2001)}]{Filippov:2001}
{Filippov}, B.~P., \& {Den}, O.~G. 2001, \jgr, 106, 25177,
  \dodoi{10.1029/2000JA004002}

\bibitem[{{Forbes}(2000)}]{Forbes:2000}
{Forbes}, T.~G. 2000, \jgr, 105, 23153, \dodoi{10.1029/2000JA000005}

\bibitem[{{Forbes} \& {Isenberg}(1991)}]{Forbes:1991}
{Forbes}, T.~G., \& {Isenberg}, P.~A. 1991, \apj, 373, 294,
  \dodoi{10.1086/170051}

\bibitem[{{Gibson} {et~al.}(2004){Gibson}, {Fan}, {Mandrini}, {Fisher}, \&
  {D\'emoulin}}]{Gibson:2004}
{Gibson}, S.~E., {Fan}, Y., {Mandrini}, C., {Fisher}, G., \& {D\'emoulin}, P.
  2004, \apj, 617, 600, \dodoi{10.1086/425294}

\bibitem[{{Gibson} {et~al.}(2006){Gibson}, {Foster}, {Burkepile}, {de Toma}, \&
  {Stanger}}]{Gibson:2006}
{Gibson}, S.~E., {Foster}, D., {Burkepile}, J., {de Toma}, G., \& {Stanger}, A.
  2006, \apj, 641, 590, \dodoi{10.1086/500446}

\bibitem[{{Gun{\'a}r} \& {Mackay}(2015)}]{Gunar:2015}
{Gun{\'a}r}, S., \& {Mackay}, D.~H. 2015, \apj, 803, 64,
  \dodoi{10.1088/0004-637X/803/2/64}

\bibitem[{{Gun{\'a}r} {et~al.}(2013){Gun{\'a}r}, {Mackay}, {Anzer}, \&
  {Heinzel}}]{Gunar:2013}
{Gun{\'a}r}, S., {Mackay}, D.~H., {Anzer}, U., \& {Heinzel}, P. 2013, \aap,
  551, A3, \dodoi{10.1051/0004-6361/201220597}

\bibitem[{{Hillier}(2018)}]{Hillier:2018}
{Hillier}, A. 2018, Reviews of Modern Plasma Physics, 2, 1,
  \dodoi{10.1007/s41614-017-0013-2}

\bibitem[{{Hillier} {et~al.}(2012){Hillier}, {Isobe}, {Shibata}, \&
  {Berger}}]{Hillier:2012}
{Hillier}, A., {Isobe}, H., {Shibata}, K., \& {Berger}, T. 2012, \apj, 756,
  110, \dodoi{10.1088/0004-637X/756/2/110}

\bibitem[{{James} {et~al.}(2018){James}, {Valori}, {Green}, {Liu}, {Cheung},
  {Guo}, \& {van Driel-Gesztelyi}}]{James:2018}
{James}, A.~W., {Valori}, G., {Green}, L.~M., {et~al.} 2018, \apjl, 855, L16,
  \dodoi{10.3847/2041-8213/aab15d}

\bibitem[{{James} {et~al.}(2017){James}, {Green}, {Palmerio}, {Valori}, {Reid},
  {Baker}, {Brooks}, {van Driel-Gesztelyi}, \& {Kilpua}}]{James:2017}
{James}, A.~W., {Green}, L.~M., {Palmerio}, E., {et~al.} 2017, \solphys, 292,
  71, \dodoi{10.1007/s11207-017-1093-4}

\bibitem[{{Jenkins} {et~al.}(2018){Jenkins}, {Long}, {van Driel-Gesztelyi}, \&
  {Carlyle}}]{Jenkins:2018}
{Jenkins}, J.~M., {Long}, D.~M., {van Driel-Gesztelyi}, L., \& {Carlyle}, J.
  2018, \solphys, 293, 7, \dodoi{10.1007/s11207-017-1224-y}

\bibitem[{{Kaiser} {et~al.}(2008){Kaiser}, {Kucera}, {Davila}, {St.~Cyr},
  {Guhathakurta}, \& {Christian}}]{Kaiser:2008}
{Kaiser}, M.~L., {Kucera}, T.~A., {Davila}, J.~M., {et~al.} 2008, \ssr, 136, 5,
  \dodoi{10.1007/s11214-007-9277-0}

\bibitem[{{Kaneko} \& {Yokoyama}(2018)}]{Kaneko:2018}
{Kaneko}, T., \& {Yokoyama}, T. 2018, \apj, 869, 136,
  \dodoi{10.3847/1538-4357/aaee6f}

\bibitem[{{Klimchuk}(2001)}]{Klimchuk:2001}
{Klimchuk}, J.~A. 2001, Washington DC American Geophysical Union Geophysical
  Monograph Series, 125, \dodoi{10.1029/GM125p0143}

\bibitem[{{Labrosse} {et~al.}(2010){Labrosse}, {Heinzel}, {Vial}, {Kucera},
  {Parenti}, {Gun{\'a}r}, {Schmieder}, \& {Kilper}}]{Labrosse:2010}
{Labrosse}, N., {Heinzel}, P., {Vial}, J.~C., {et~al.} 2010, \ssr, 151, 243,
  \dodoi{10.1007/s11214-010-9630-6}

\bibitem[{{Lemen} {et~al.}(2012){Lemen}, {Title}, {Akin}, {Boerner}, {Chou}, \&
  {Drake}}]{Lemen:2012}
{Lemen}, J.~R., {Title}, A.~M., {Akin}, D.~J., {et~al.} 2012, \solphys, 275,
  17, \dodoi{10.1007/s11207-011-9776-8}

\bibitem[{{Lin} {et~al.}(2002){Lin}, {van Ballegooijen}, \&
  {Forbes}}]{Lin:2002}
{Lin}, J., {van Ballegooijen}, A.~A., \& {Forbes}, T.~G. 2002, \jgr (Space
  Physics), 107, 1438, \dodoi{10.1029/2002JA009486}

\bibitem[{{Low}(1981)}]{Low:1981}
{Low}, B.~C. 1981, \apj, 246, 538, \dodoi{10.1086/158954}

\bibitem[{{Low}(1999)}]{Low:1999}
{Low}, B.~C. 1999, in American Institute of Physics Conference Series, Vol.
  471, 109--114

\bibitem[{{Low} {et~al.}(2003){Low}, {Fong}, \& {Fan}}]{Low:2003}
{Low}, B.~C., {Fong}, B., \& {Fan}, Y. 2003, \apj, 594, 1060,
  \dodoi{10.1086/377042}

\bibitem[{{Lynch} {et~al.}(2004){Lynch}, {Antiochos}, {MacNeice}, {Zurbuchen},
  \& {Fisk}}]{Lynch:2004}
{Lynch}, B.~J., {Antiochos}, S.~K., {MacNeice}, P.~J., {Zurbuchen}, T.~H., \&
  {Fisk}, L.~A. 2004, \apj, 617, 589, \dodoi{10.1086/424564}

\bibitem[{{Martens} \& {Kuin}(1989)}]{Martens:1989}
{Martens}, P.~C.~H., \& {Kuin}, N.~P.~M. 1989, \solphys, 122, 263,
  \dodoi{10.1007/BF00912996}

\bibitem[{{Martin}(1998)}]{Martin:1998}
{Martin}, S.~F. 1998, \solphys, 182, 107, \dodoi{10.1023/A:1005026814076}

\bibitem[{{McCauley} {et~al.}(2015){McCauley}, {Su}, {Schanche}, {Evans}, {Su},
  {McKillop}, \& {Reeves}}]{McCauley:2015}
{McCauley}, P.~I., {Su}, Y.~N., {Schanche}, N., {et~al.} 2015, \solphys, 290,
  1703, \dodoi{10.1007/s11207-015-0699-7}

\bibitem[{{Moore} {et~al.}(2001){Moore}, {Sterling}, {Hudson}, \&
  {Lemen}}]{Moore:2001}
{Moore}, R.~L., {Sterling}, A.~C., {Hudson}, H.~S., \& {Lemen}, J.~R. 2001,
  \apj, 552, 833, \dodoi{10.1086/320559}

\bibitem[{{Palmerio} {et~al.}(2017){Palmerio}, {Kilpua}, {James}, {Green},
  {Pomoell}, {Isavnin}, \& {Valori}}]{Palmerio:2017}
{Palmerio}, E., {Kilpua}, E.~K.~J., {James}, A.~W., {et~al.} 2017, \solphys,
  292, 39, \dodoi{10.1007/s11207-017-1063-x}

\bibitem[{{Pesnell} {et~al.}(2012){Pesnell}, {Thompson}, \&
  {Chamberlin}}]{Pesnell:2012}
{Pesnell}, W.~D., {Thompson}, B.~J., \& {Chamberlin}, P.~C. 2012, \solphys,
  275, 3, \dodoi{10.1007/s11207-011-9841-3}

\bibitem[{{Petrie} {et~al.}(2007){Petrie}, {Blokland}, \&
  {Keppens}}]{Petrie:2007}
{Petrie}, G.~J.~D., {Blokland}, J.~W.~S., \& {Keppens}, R. 2007, \apj, 665,
  830, \dodoi{10.1086/519276}

\bibitem[{{Priest} {et~al.}(1989){Priest}, {Hood}, \& {Anzer}}]{Priest:1989}
{Priest}, E.~R., {Hood}, A.~W., \& {Anzer}, U. 1989, \apj, 344, 1010,
  \dodoi{10.1086/167868}

\bibitem[{{Reeves} \& {Forbes}(2005)}]{Reeves:2005}
{Reeves}, K.~K., \& {Forbes}, T.~G. 2005, in Coronal and Stellar Mass
  Ejections, IAU Symposium Proceedings of the International Astronomical Union
  226, Held 13-17 September, Beijing, edited by K. Dere, J. Wang, and Y. Yan.
  Cambridge: Cambridge University Press, 2005., pp.250-255, Vol. 226, 250--255

\bibitem[{{R{\'e}gnier} {et~al.}(2011){R{\'e}gnier}, {Walsh}, \&
  {Alexander}}]{Regnier:2011}
{R{\'e}gnier}, S., {Walsh}, R.~W., \& {Alexander}, C.~E. 2011, \aap, 533, L1,
  \dodoi{10.1051/0004-6361/201117381}

\bibitem[{{Reva} {et~al.}(2017){Reva}, {Kirichenko}, {Ulyanov}, \&
  {Kuzin}}]{Reva:2017}
{Reva}, A.~A., {Kirichenko}, A.~S., {Ulyanov}, A.~S., \& {Kuzin}, S.~V. 2017,
  \apj, 851, 108, \dodoi{10.3847/1538-4357/aa9986}

\bibitem[{{Rompolt}(1990)}]{Rompolt:1990}
{Rompolt}, B. 1990, Hvar Observatory Bulletin, 14, 37

\bibitem[{{Rust}(2003)}]{Rust:2003}
{Rust}, D.~M. 2003, Advances in Space Research, 32, 1895,
  \dodoi{10.1016/S0273-1177(03)90623-5}

\bibitem[{{Scherrer} {et~al.}(1995){Scherrer}, {Bogart}, {Bush}, {Hoeksema},
  {Kosovichev}, {Schou}, {Rosenberg}, {Springer}, {Tarbell}, {Title},
  {Wolfson}, {Zayer}, \& {MDI Engineering Team}}]{Scherrer:1995}
{Scherrer}, P.~H., {Bogart}, R.~S., {Bush}, R.~I., {et~al.} 1995, \solphys,
  162, 129, \dodoi{10.1007/BF00733429}

\bibitem[{{Schmahl} \& {Hildner}(1977)}]{Schmahl:1977}
{Schmahl}, E., \& {Hildner}, E. 1977, \solphys, 55, 473,
  \dodoi{10.1007/BF00152588}

\bibitem[{{Schutgens} \& {T{\'o}th}(1999)}]{Schutgens:1999}
{Schutgens}, N.~A.~J., \& {T{\'o}th}, G. 1999, \aap, 345, 1038.
\newblock \doarXiv{astro-ph/9903128}

\bibitem[{{Seaton} {et~al.}(2011){Seaton}, {Mierla}, {Berghmans}, {Zhukov}, \&
  {Dolla}}]{Seaton:2011}
{Seaton}, D.~B., {Mierla}, M., {Berghmans}, D., {Zhukov}, A.~N., \& {Dolla}, L.
  2011, \apjl, 727, L10, \dodoi{10.1088/2041-8205/727/1/L10}

\bibitem[{{Titov} \& {D{\'e}moulin}(1999)}]{Titov:1999}
{Titov}, V.~S., \& {D{\'e}moulin}, P. 1999, \aap, 351, 707

\bibitem[{{T{\"o}r{\"o}k} \& {Kliem}(2005)}]{Torok:2005}
{T{\"o}r{\"o}k}, T., \& {Kliem}, B. 2005, \apjl, 630, L97,
  \dodoi{10.1086/462412}

\bibitem[{{van Ballegooijen} \& {Martens}(1989)}]{vanballegooijen:1989}
{van Ballegooijen}, A.~A., \& {Martens}, P.~C.~H. 1989, \apj, 343, 971,
  \dodoi{10.1086/167766}

\bibitem[{{van Driel-Gesztelyi} {et~al.}(2003){van Driel-Gesztelyi},
  {D{\'e}moulin}, {Mandrini}, {Harra}, \& {Klimchuk}}]{vanDriel:2003}
{van Driel-Gesztelyi}, L., {D{\'e}moulin}, P., {Mandrini}, C.~H., {Harra}, L.,
  \& {Klimchuk}, J.~A. 2003, \apj, 586, 579, \dodoi{10.1086/367633}

\bibitem[{{van Tend} \& {Kuperus}(1978)}]{vanTend:1978}
{van Tend}, W., \& {Kuperus}, M. 1978, \solphys, 59, 115,
  \dodoi{10.1007/BF00154935}

\bibitem[{{Williams} {et~al.}(2013){Williams}, {Baker}, \& {van
  Driel-Gesztelyi}}]{Williams:2013}
{Williams}, D.~R., {Baker}, D., \& {van Driel-Gesztelyi}, L. 2013, \apj, 764,
  165, \dodoi{10.1088/0004-637X/764/2/165}

\bibitem[{{Wuelser} {et~al.}(2004){Wuelser}, {Lemen}, {Tarbell}, {Wolfson},
  {Cannon}, \& {Carpenter}}]{Wuelser:2004}
{Wuelser}, J.-P., {Lemen}, J.~R., {Tarbell}, T.~D., {et~al.} 2004, in
  \procspie, Vol. 5171, Telescopes and Instrumentation for Solar Astrophysics,
  ed. S.~{Fineschi} \& M.~A. {Gummin}, 111--122

\bibitem[{{Xia} {et~al.}(2012){Xia}, {Chen}, \& {Keppens}}]{Xia:2012}
{Xia}, C., {Chen}, P.~F., \& {Keppens}, R. 2012, \apj, 748, L26,
  \dodoi{10.1088/2041-8205/748/2/L26}

\bibitem[{{Xia} \& {Keppens}(2016)}]{Xia:2016}
{Xia}, C., \& {Keppens}, R. 2016, \apj, 823, 22,
  \dodoi{10.3847/0004-637X/823/1/22}

\bibitem[{{Zhou} {et~al.}(2018){Zhou}, {Xia}, {Keppens}, {Fang}, \&
  {Chen}}]{Zhou:2018}
{Zhou}, Y.-H., {Xia}, C., {Keppens}, R., {Fang}, C., \& {Chen}, P.~F. 2018,
  \apj, 856, 179, \dodoi{10.3847/1538-4357/aab614}

\bibitem[{{Zuccarello} {et~al.}(2016){Zuccarello}, {Aulanier}, \&
  {Gilchrist}}]{Zuccarello:2016}
{Zuccarello}, F.~P., {Aulanier}, G., \& {Gilchrist}, S.~A. 2016, \apjl, 821,
  L23, \dodoi{10.3847/2041-8205/821/2/L23}

\bibitem[{{Zuccarello} {et~al.}(2014){Zuccarello}, {Seaton}, {Mierla},
  {Poedts}, {Rachmeler}, {Romano}, \& {Zuccarello}}]{Zuccarello:2014}
{Zuccarello}, F.~P., {Seaton}, D.~B., {Mierla}, M., {et~al.} 2014, \apj, 785,
  88, \dodoi{10.1088/0004-637X/785/2/88}

\end{thebibliography}

\end{document}